\documentclass[11pt,a4paper]{article}
\usepackage{jcappub}

\usepackage{graphicx}
\usepackage{dcolumn}
\usepackage{bm}
\usepackage{relsize} 
\usepackage{multirow} 

\usepackage{rotating} 
\usepackage{makecell} 

\hypersetup{colorlinks=true, urlcolor=blue, citecolor=blue}
    
\newcommand{\HeitlerMatthews}{1937Carlson, 1954Heitler, 2005Matthews}
\newcommand{\CONEX}{CONEX1, CONEX2}
\newcommand{\CONEXxCorsika}{CONEXxCorsika1, CONEXxCorsika2}
\newcommand{\QGSJetII}{2011QGSJetII-04, 2013QGSJetII-04}
\newcommand{\UrQMD}{UrQMD1, UrQMD2}

\begin{document}

\title{Modeling Strangeness Enhancements to resolve the Muon Excess\\ in Cosmic Ray Extensive Air Shower Data}

\author{Julien Manshanden,}
\emailAdd{julien.manshanden@desy.de}
\author{Günter Sigl}
\emailAdd{guenter.sigl@desy.de}
\author{and Maria V. Garzelli}
\emailAdd{maria.vittoria.garzelli@desy.de}
\affiliation{%
 II. Institut für Theoretische Physik, Universität Hamburg,\\ Luruper Chaussee 149, 22761 Hamburg, Germany
}%


\date{\today}

\abstract{
    Experimental observations of extensive air showers have revealed an excess of the muon content with respect to their theoretical simulations, which we refer to as the muon puzzle. This muon puzzle hampers a precise determination of the ultra-high-energy cosmic ray mass composition. We investigate the potential of producing states of dense quark-gluon matter (which we call fireballs) to resolve the muon puzzle as quantified 
    with data from the Pierre Auger Observatory on the depth of the shower maximum and the number of muons at ground.
    Adopting a phenomenological fireball model, we find that the inelasticity enhancement associated with the formation of a plasma state 
    is in tension with data on the electromagnetic longitudinal shower development. Instead, we restrict the fireball model to only enhance the strangeness produced in 
    Standard Model hadronic interactions, {and dub this model the strangeball model}. 
    With an analytic approach based on the Heitler-Matthews model we then find explicit sets of {strangeball} parameters that resolve the muon puzzle. Constraints from data on shower-to-shower fluctuations of the muon number require strangeness enhancements already at energies accessible to current-generation collider experiments. At Tevatron and LHC energies we estimate 40\% of the interactions to produce {strangeballs}, corresponding to a $5-9$\% increase of the {average fraction of energy retained in the hadronic cascade} compared to predictions from current hadronic interaction models. A comparison with relevant measurements of the LHCf and LHCb detectors does not directly exclude this scenario, 
    {though the obtained tension with LHCb suggests a stringent test at 14 TeV}.
}

\arxivnumber{2208.04266}

\maketitle


\section{Introduction}

Cosmic rays (CRs) have been measured to attain energies exceeding $10^{20}$ eV \cite{AugerSpectrum}. These energies are indicative of 
the extreme environments responsible for their acceleration. Gaining an understanding of the CR origin is complicated by the non-trivial propagation to Earth, with the charged nature of CRs playing a prominent role. Precise measurements of the CR mass composition  
{should} provide stringent tests for current astrophysical scenarios. Due to the rapidly decreasing CR flux, the masses of CRs with energies above $10^{15}$ eV cannot be measured directly and must instead be inferred from the induced air showers, which leave characteristic imprints in the atmosphere and on the ground.

Some of these imprints are sensitive to the CR mass. In particular, the depth at which the shower reaches its maximum $X_\mathrm{max}$ is a traditional mass indicator (e.g., \cite{2014AugerXmaxmeasurement, 2014AugerXmaxcomp}). A consistent picture among further independent mass indicators 
could enhance the credibility and accuracy of the inferred composition (e.g., \cite{2016AugerMixed,2017AugerInferencesTests}). Instead, various experiments found that the number of muons reaching the ground points towards a significantly heavier composition, 
or inversely, that current air shower simulations underestimate muon production \cite{2019Dembinski, 2021MuonPuzzle, 2021WHISP}. This is known as the muon puzzle and can be traced back to an incomplete understanding of the hadronic interactions in air showers. Therefore, the muon puzzle also constitutes an opportunity for cosmic ray experiments to provide predictions for collider experiments regarding potentially new physics. 

With ad-hoc adjustments to hadronic interaction parameters, the number of muons was found \cite{2010Ulrich} to mainly depend on the total multiplicity and the fraction of all pions that are neutral. 
Considering also the impact of these adjustments on $X_\mathrm{max}$, 
it was shown \cite{2020Baur} that an appropriate suppression of the fraction of energy going into electromagnetic (EM) particles has the potential to resolve the muon puzzle. Most recently this was also shown to be the case \cite{2022AnchordoquiPiKswap} when considering a phenomenological procedure of swapping pions and kaons.

Here\footnote{A more extensive description of the studies of sections~\ref{sec:impactEAS} - \ref{sec:ApplyAuger} can be found in ref.~\cite{JulienThesis}.} we adopt a similar procedure to investigate whether producing fireballs \cite{AnchordoquiFireball} --- one of the various models proposed in the literature \cite{2020Baur, AnchordoquiFireball, 2012Muniz, 2013Farrar, 2019PierogICRC, 2021PierogICRC} --- in the first few interactions of an air shower could equally solve the muon puzzle. We subsequently attempt to interpret data from the Pierre Auger Observatory \cite{2015AugerDetails} in terms of {macroscopic} hadronic interaction properties by extending the analytic formalism known as the Heitler-Matthews model \cite{\HeitlerMatthews}. Finally, we compare our results to measurements from the LHCf and LHCb detectors {at the Large Hadron Collider (LHC)} and suggest some follow-up research to drive this synergy between cosmic ray and collider experiments.

This paper is structured as follows. In sections~\ref{sec:fireballtheory} and~\ref{sec:impactEAS} we describe the fireball model and study its impact on the relevant air shower observables, respectively. The extension to the Heitler-Matthews model is outlined in section~\ref{sec:HeitlerMatthews}, and subsequently applied to Auger data in section~\ref{sec:ApplyAuger}. We discuss some implications of these results for measurements at the LHC in section~\ref{sec:LHCimplications}, and draw our conclusions in section~\ref{sec:conclusions}.




\section{The Fireball Model}
\label{sec:fireballtheory}

A fireball state of matter is hypothesized \cite{AnchordoquiFireball} to form when the energy density in a collision exceeds some threshold value, estimated {as} 1 GeV/fm$^{3}$.  
Upon formation, the fireball is a plasma consisting of deconfined up and down quarks and gluons maintained in both kinetic and chemical equilibrium. An associated high baryochemical potential leads to the fragmentation of gluons into strange quarks, resulting in an enhanced production of strange secondaries upon hadronization. This indirectly suppresses the neutral pion production compared to the Standard Model case, 
altering the air shower evolution \cite{AnchordoquiFireball}.

{In the present study we do not further develop this microscopic treatment, nor trace its connection to thermodynamical descriptions of the quark-gluon plasma (QGP) \cite{2009Andronic, 2016Braun-Munzinger,2016Dusling,2017Koch,2018Busza,2020Gazdzicki,2022ALICEQCDJourney}. While the details regarding such a connection can be found in ref.~\cite{AnchordoquiFireball}, we proceed here with a phenomenological approach, 
only considering an effective enhancement of the strange particle content and an altered multiplicity and elasticity associated with fireball interactions. We note, however, that strangeness enhancements are in fact observed in heavy-ion collisions \cite{2005BRAHMS, 2005PHENIX,2005STAR,2008Becattini} and more recently, with ALICE at the LHC also in proton-lead \cite{2014ALICEpPb,2016ALICEpPb} and proton-proton collisions \cite{2017ALICEStrangenesspp}. In contrast to a previous study, considering the production of QGPs only in the first interaction of an air shower \cite{2017LaHurd}, the ALICE observations open up the possibility of copious production of such quark-matter states -- perhaps this fireball -- in air showers and thereby potentially contribute to a solution to the muon puzzle \cite{2020Baur,2021MuonPuzzle}.}




In the following we take our reference frame to be that of the Earth, with the atmospheric particles at rest providing the fixed target for the energetic air shower projectiles. These projectiles consist of the primary CRs and the produced secondaries.

The fireball 
state was proposed to be modeled \cite{AnchordoquiProceeding} from Standard Model interactions in two steps. The enhanced multiplicity and inelasticity associated with a plasma state can be mimicked by repeated \emph{in situ} collisions of the projectile (and its secondaries) with air nuclei until a specific condition is satisfied. We take this condition to be that secondary nucleons or nuclear fragments only participate in this process if their energy {$E_\mathrm{frag}$} is above a fixed fraction $f_\mathrm{thres}$
of the energy of the projectile $E_\mathrm{proj}$ that initiates the fireball\footnote{Under the simplification of only nuclear secondaries, our $f_\mathrm{thres}$-parameter introduces a minimum multiplicity for fireball interactions: $n_\mathrm{mult} \geq 1/f_\mathrm{thres}$, with an equality symbol when equally distributing energy {between} the secondaries.}{:
\begin{align}
    E_\mathrm{frag} > f_\mathrm{thres} \cdot E_\mathrm{proj} \;.
\end{align}}
The enhanced production of strange particles and the associated suppression of the neutral pion production is then mimicked by swapping all pions and kaons while conserving energy, direction of momentum, and charge (with an equal probability of changing a $\pi^0$ into a $K^0_L$ or a $K^0_S$). 

For this phenomenological model it is further necessary to define the condition of producing a fireball. For a fixed projectile energy, an energy-density threshold as stated above translates to a fireball-production probability due to variations of the impact parameter. We expect this probability $p(E)$ to grow with the projectile energy{, as more peripheral collisions can potentially attain the fireball state. This is closely related to the core-corona picture, which was successfully applied in interpreting the centrality dependence of various observables in heavy-ion collisions \cite{2005Bozek,2007Werner,2008Aichelin,2009Becattini}. The central region of an interaction, the core, produces a new quark matter state (similar to our fireball), while the outer region, the corona, produces particles through standard string fragmentation. From geometrical considerations and in agreement with these measurements the contribution of the core was found to increase with centrality, see, e.g., figure 4 of ref.~\cite{2009Becattini}. Given the black disk limit of protons \cite{2015Block}, a similar increase up to unity would then be expected with rising center of mass energy. Whereas in the core-corona model one typically has a superposition of both kinds of interactions in one event, the fireball is only produced with a certain probability in a given interaction.
}

We parametrize {this fireball-production probability} as follows:
\begin{align}
    p(E) =
\begin{cases}
0\;, &\mathrm{if} \; E<E_\mathrm{min}, \\[0.5em]
\mathlarger{ \left( \frac{\log(E/E_\mathrm{min})}{\log(E_\mathrm{max}/E_\mathrm{min})} \right)^n}\;, \qquad \qquad &\mathrm{if} \; E_\mathrm{min}<E<E_\mathrm{max}, \\[1em]
1\;, &\mathrm{if} \; E>E_\mathrm{max}.
\end{cases}
\label{eq:probfireball}
\end{align}
In this way no fireballs are produced below some minimum energy $E_\mathrm{min}$. Then the production probability grows logarithmically 
(for $n=1$) up to 
some maximum energy $E_\mathrm{max}$, above which every interaction produces  
a fireball state.
{The approximation of having a logarithmic growth is in line with related previous works, see refs.~\cite{2013Farrar, 2020Baur}.}


{Note that swapping all pions and kaons effectively dictates the size of the strangeness enhancement in fireball interactions. This corresponds to the first approximation of ref.~\cite{AnchordoquiProceeding}, and should also be regarded as such in this study. More flexibility could be incorporated by swapping a variable fraction of the pions and kaons. However, in the following analysis this is degenerate with the probability of eq.~\ref{eq:probfireball}, and thus would not alter our results. The power $n$ could absorb a potential energy dependence of the size of the strangeness enhancement from a single fireball, though a detailed assessment of this size is beyond the scope of this paper.}

To summarize, our implementation of the fireball model has four parameters: one ($f_\mathrm{thres}$) controlling the plasma state, and three ($E_\mathrm{min}$, $E_\mathrm{max}$, $n$) regulating the fireball-production probability.

\section{The Impact of Fireballs on Air Shower Observables}
\label{sec:impactEAS}

\subsection{Method}

To study the effect of the fireball model on the development of air showers we implemented our phenomenological model into the \textsc{Conex} (version 7.5) \cite{\CONEX} module \cite{\CONEXxCorsika} of \textsc{Corsika} (version 7.74) \cite{Corsika}. 
This implementation constituted altering 
the \textsc{cnexus} subroutine, which functions as the interface between the shower evolution and the hadronic interaction models. We compute the longitudinal (one-dimensional) shower evolution with the \textsc{cascade TTT} and \textsc{AugerHit} options, and extract $X_\mathrm{max}$ and the number of muons $N_\mu$ at an altitude of 1425~m.  
We further set the muon detection threshold to 0.3 GeV and the shower inclination to $\theta=67^\circ$ in accordance with the Pierre Auger Observatory study of inclined showers \cite{2015AugerMuons}. With this set-up the first part of the shower 
is computed with a Monte Carlo simulation, enabling us to also study shower-to-shower fluctuations.

We consider data from the Pierre Auger Observatory (hereafter just `Auger data') on the average $\langle . \rangle$ and fluctuations $\sigma(.)$ of $X_\mathrm{max}$ and $R_\mu = N_\mu/1.455\cdot10^7$ (for $\theta=67^\circ$) as presented at the International Cosmic Ray Conference (ICRC) in 2019 \cite{2019ICRC}. At this stage we want to investigate whether there are fireball settings with which we can obtain a consistent interpretation of the data in terms of the CR mass composition. 
To do so, we explore the fireball parameter space by first fixing $f_\mathrm{thres}=0.01$ (invoking a plasma) 
and $E_\mathrm{min}=10^{15}$ eV ($\Leftrightarrow \sqrt{s}_\mathrm{min}\approx 1.4$ TeV), and then we sample $E_\mathrm{max} \in \{10^{17}, 10^{18}, 10^{19}, 10^{20}\}$ eV and $n \in \{1, 2, 4, 8, 1000\}$. Note that with $n=1000$, $p(E)$ represents a step function at $E_\mathrm{max}$.

For each of these fireball settings we simulate 2100 showers from proton, helium, nitrogen, silicon and iron CR primaries at an energy of 10 EeV. In this section we do not consider further energies and thus focus on the corresponding data points at this energy. Regarding the hadronic interaction models, we use \textsc{QGSJetII-04} \cite{\QGSJetII}, EPOS-LHC \cite{EPOSLHC}, and \textsc{Sibyll-2.3d} \cite{2020Sibyll23d_PRD} at high energies, and \textsc{UrQMD} \cite{\UrQMD} at low energies.

\subsection{Results}

\begin{figure}[h]
    \makebox[1\textwidth]{
        \includegraphics[width=1.12\textwidth]{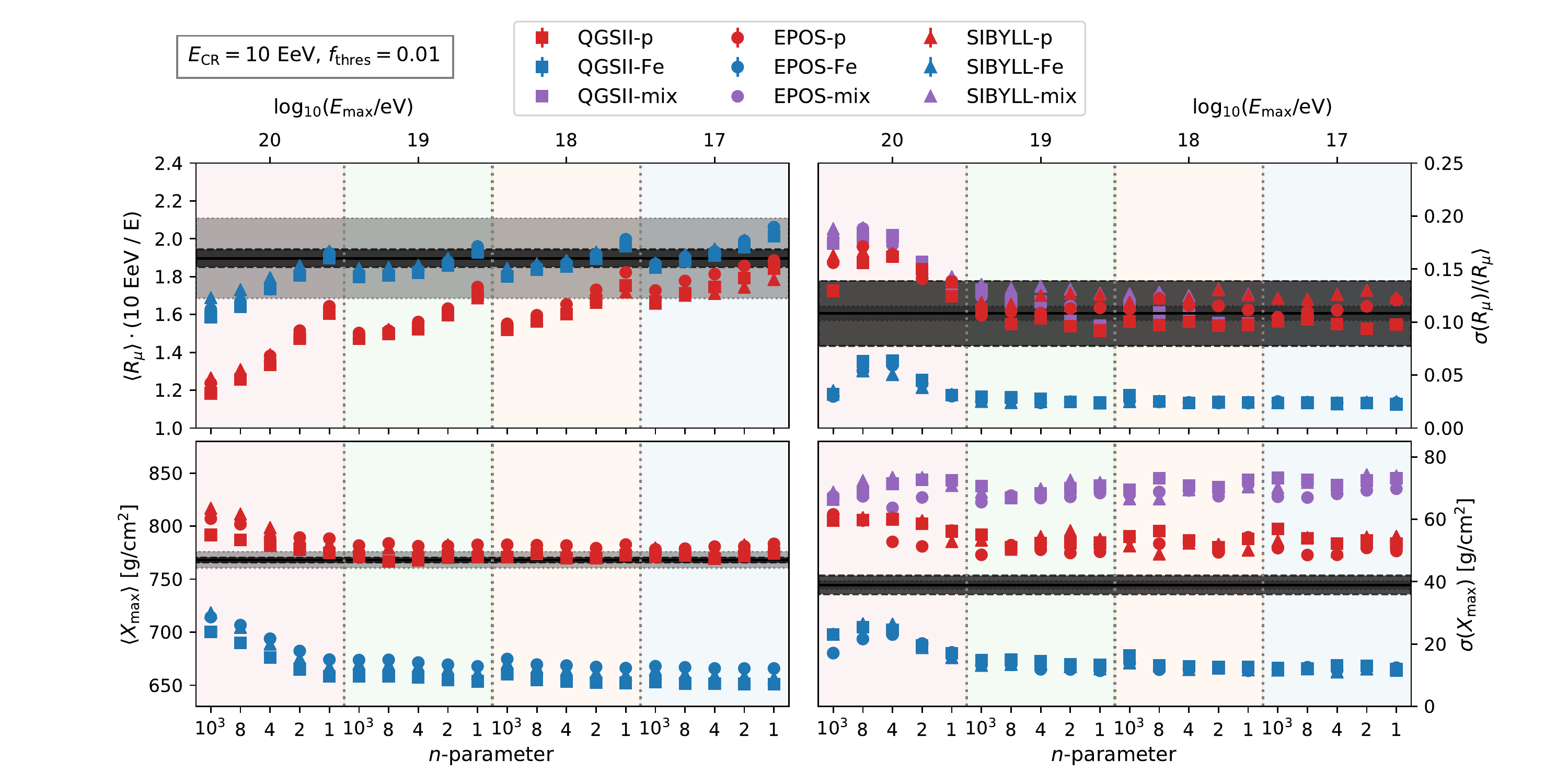}
    }
    \caption{\label{fig:obsfstop2}Impact of various settings of the fireball model on the ranges of $\langle R_\mu \rangle$ (top-left), $\sigma(R_\mu)/\langle R_\mu \rangle$ (top-right), $\langle X_\mathrm{max} \rangle$ (bottom-left) and $\sigma(X_\mathrm{max})$ (bottom-right) predictions from 10 EeV air showers. Varied on the $x$-axis are $E_\mathrm{max}$ (top axis) and $n$ (bottom axis), while $E_\mathrm{min}=10^{15}$ eV and $f_\mathrm{thres}=0.01$ are fixed. Showers are simulated using the high-energy hadronic interaction models \textsc{QGSJetII-04} (squares), \textsc{EPOS-LHC} (circles), and \textsc{Sibyll-2.3d} (triangles) in combination with a pure proton (red) and pure iron (blue) composition. Further simulating helium, nitrogen and silicon showers enabled the computation of the mixed compositions that maximize the fluctuations (purple). Data (black horizontal lines) with systematic (light gray bands) and statistical (dark gray bands) uncertainties are from the Pierre Auger Observatory as presented at the ICRC 2019 \cite{2019ICRC}.}
\end{figure}

The minimum and maximum values of the observables $\langle X_\mathrm{max} \rangle$, $\sigma(X_\mathrm{max})$, $\langle R_\mu \rangle$, and $\sigma(R_\mu)/\langle R_\mu \rangle$ under variations of the mass composition are shown in figure~\ref{fig:obsfstop2}, with the fireball settings varied along the $x$-axis. For the average observables 
these extremes simply correspond to the proton and iron predictions, but, in case of fluctuations, a mixed composition may lead to even larger values. Note that the left-most setting ($E_\mathrm{max} = 10^{20}$ eV and $n=1000$) introduces fireballs above the energy of the primary CR and thus corresponds to the Standard Model. Accordingly, comparisons with the left-most points reveal the impact of the fireball model on the air shower observables.

For some settings the fireball model sufficiently increases $\langle R_\mu \rangle$ such that the extremes encompass the Auger data point (horizontal line). Simultaneously, the difference between these extremes decreases as fireballs become more abundant at lower energies. This is a universal feature of models invoking a mass-independent increase of the muon number,  
ultimately making the muon number less sensitive to the mass composition.


This implementation of the fireball model also affects the other observables. The reduction of $\langle X_\mathrm{max} \rangle$ can be attributed to the production of a 
plasma and the associated enhanced inelasticity, {accelerating} the shower development. This effect seems to saturate as most energy is deposited in the EM component in the first few interactions. The relative muon fluctuations are determined by the first interaction \cite{1960Fukui,2018Cazon} and are slightly enhanced for $E_\mathrm{max}=10^{20}$ eV, $2 \leq n \leq 8$ due to a mixture of fireball and Standard Model first interactions. With only fireballs as first interactions, $\sigma(R_\mu)/\langle R_\mu \rangle$ decreases and saturates at a constant value, seeming to enforce a proton-dominated composition. A similar enhancement, decrease and saturation can be seen for $\sigma(X_\mathrm{max})$, most prominently from iron showers. In contrast to the muon observables, a splitting emerges between $\sigma(X_\mathrm{max})$ from mixed and pure proton compositions, reflecting the more pronounced separation of the proton and iron $X_\mathrm{max}$ distributions.

Once the predictions encompass the data points, we can infer the indicated mass composition $\{f_i\}$ from, e.g., $\langle X_\mathrm{max} \rangle_\mathrm{data} = \sum_i f_i \langle X_\mathrm{max} \rangle_i$, with $i$ referring to the chemical elements. Instead of solving these equations explicitly, by simultaneously considering another observable, e.g., $\langle R_\mu \rangle = \sum_i f_i \langle R_\mu \rangle_i$, we can compute the range of muon numbers that correspond to the $\langle X_\mathrm{max} \rangle$ data point under variations of the composition. For average observables, 
the extremes of this range is guaranteed to correspond to a superposition of at most two components, which can thus be readily computed.

By comparing data on different observables, this conversion method enables us to interpret the consistency of the mass composition within a specific model. In particular, we are able to convert data on $\langle X_\mathrm{max} \rangle$ to $\langle R_\mu \rangle$ and $\sigma(X_\mathrm{max}) = [\langle X_\mathrm{max}^2 \rangle - \langle X_\mathrm{max} \rangle_\mathrm{data}^2 ]^{1/2}$, and data on $\langle R_\mu \rangle$ to $\langle X_\mathrm{max} \rangle$ and $\sigma(R_\mu)/\langle R_\mu \rangle = [\langle R_\mu^2 \rangle / \langle R_\mu \rangle_\mathrm{data}^2 - 1]^{1/2}$.


\begin{figure}[h]
    \makebox[1\textwidth]{
        \includegraphics[width=1.12\textwidth]{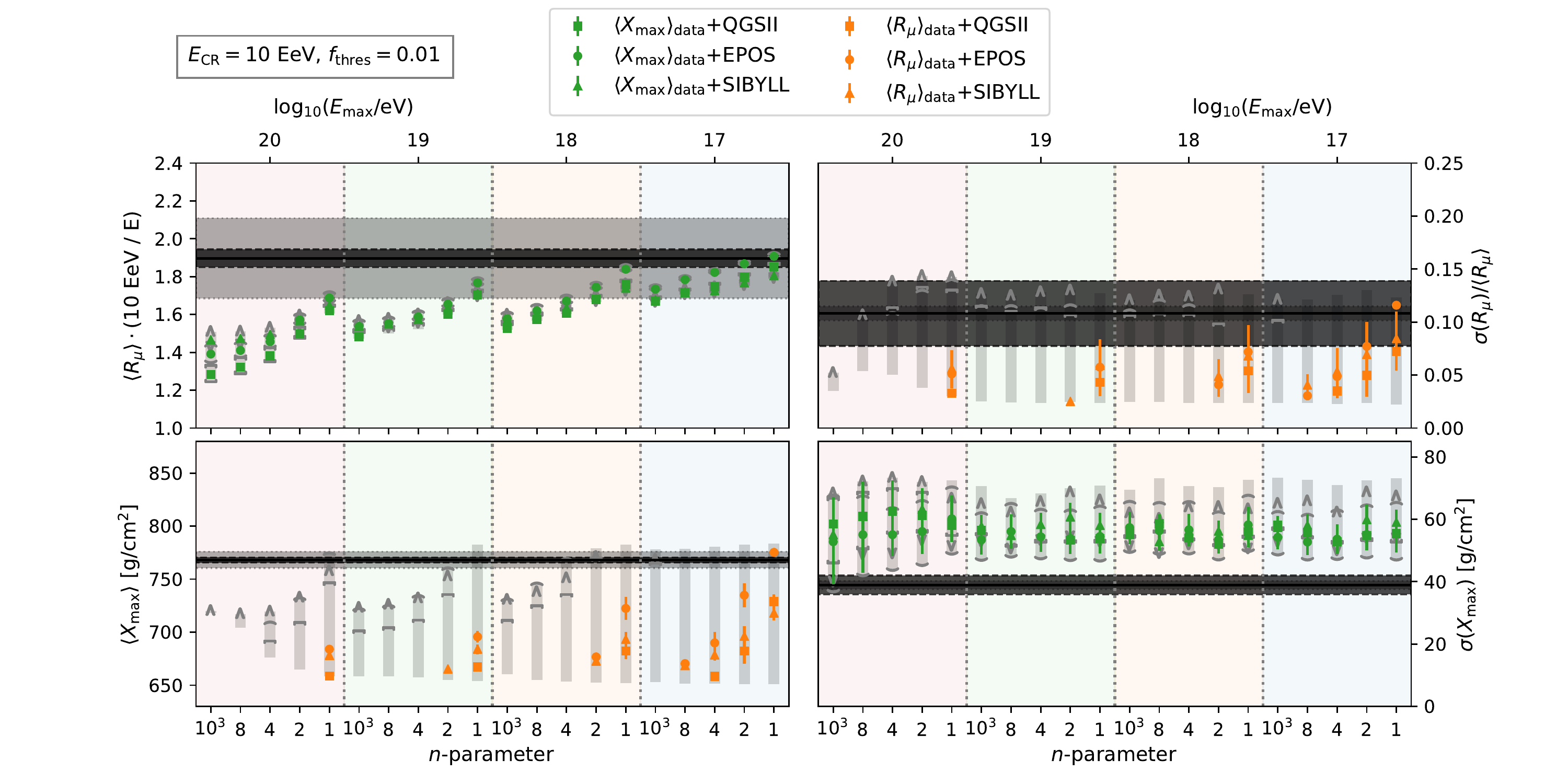}
    }
    \caption{\label{fig:mapfstop2}Similar to figure~\ref{fig:obsfstop2}, but with data on $\langle X_\mathrm{max}\rangle$ and $\langle R_\mu \rangle$ converted to the other observables through model predictions, see text for more detail. The error bars represent the variation due to a degeneracy of the composition. The (total) uncertainties are visualized through the vertical gray bars, ending where the observable falls outside of allowed range of predictions (i.e., beyond proton, iron, extreme mix).}
\end{figure}

\begin{figure}[h]
    \makebox[1\textwidth]{
        \includegraphics[width=1.12\textwidth]{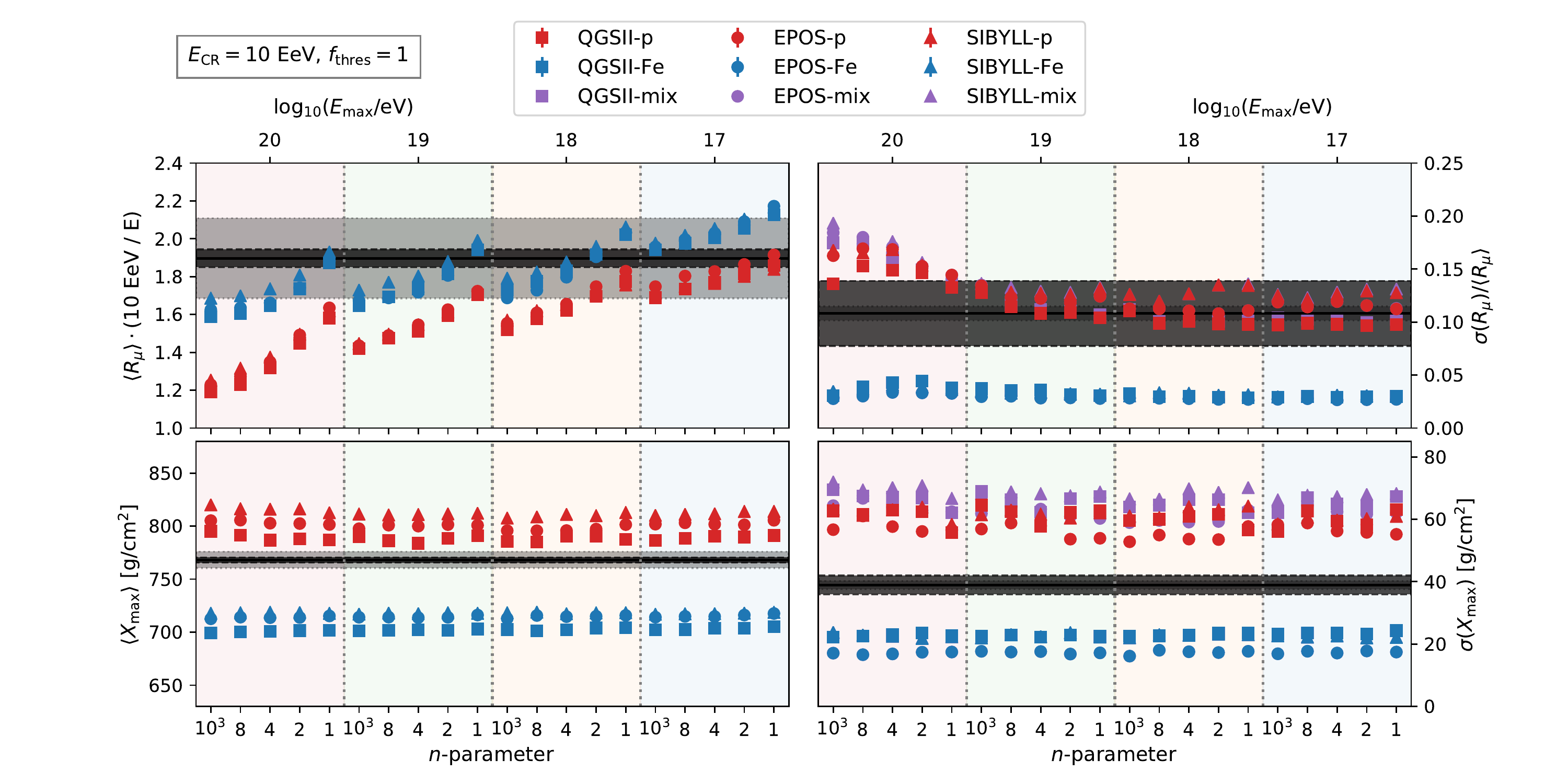}
    }
    \makebox[1\textwidth]{
        \includegraphics[width=1.12\textwidth]{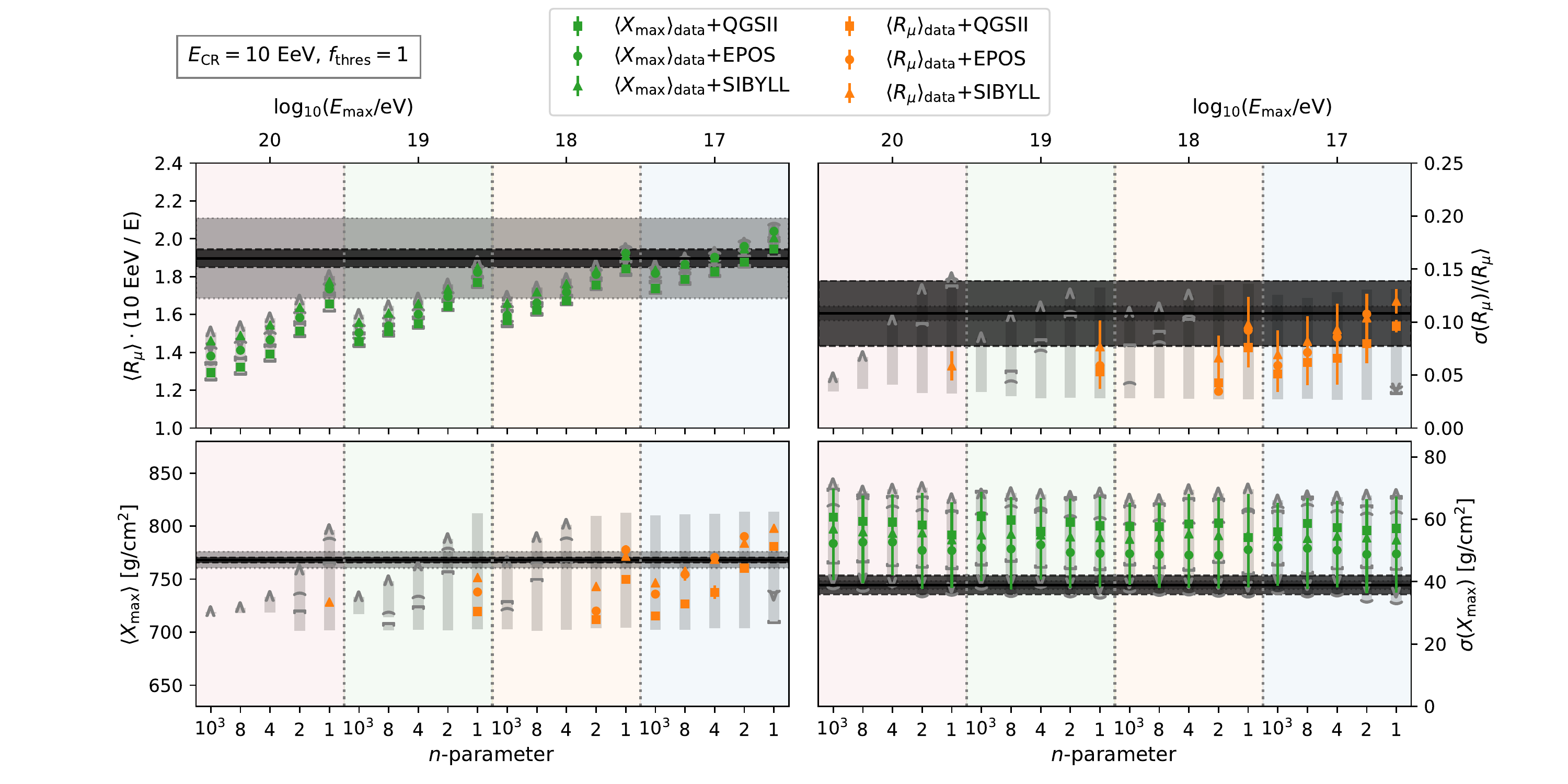}
    }
    \caption{\label{fig:obsmapfstop0} Same as figures~\ref{fig:obsfstop2} (top) and~\ref{fig:mapfstop2} (bottom), but then for $f_\mathrm{thres}=1$.}
\end{figure}

The application of this method to the fireball settings of figure~\ref{fig:obsfstop2} is shown in figure~\ref{fig:mapfstop2}. Here it becomes clear that consistency between data on $\langle X_\mathrm{max} \rangle$ and $\langle R_\mu \rangle$ is found for $E_\mathrm{max}=10^{17}$ eV and $n=1$ with \textsc{EPOS-LHC}. However, the fireball model seems to introduce an inconsistency between data on $\langle X_\mathrm{max} \rangle$ and $\sigma(X_\mathrm{max})$, with the average indicating a significantly lighter composition than the fluctuations. This can be traced back to the formation of a plasma accelerating the shower development. Therefore, we repeated the analysis for other values of $f_\mathrm{thres} \in \{1, 0.1, 0.001\}$. It turns out that the aforementioned inconsistency can only be avoided if one turns off the formation of a plasma by setting $f_\mathrm{thres}=1$, reducing fireball interactions to correspond to only swapping pions and kaons. The impact on the observables are shown in figure~\ref{fig:obsmapfstop0}. Notice that these fireballs only affect the muon number (both average and fluctuations), leaving $X_\mathrm{max}$ unaffected. For the settings summarized in table~\ref{tab:potentialsolutions} we find that a solution of the muon puzzle is possible at $E_\mathrm{CR} = 10$ EeV.

\begin{table}[h]
\centering
\caption{\label{tab:potentialsolutions} Settings of the fireball model enabling a consistent composition interpretation of data from the Pierre Auger Observatory as presented at the ICRC 2019 \cite{2019ICRC}, on both moments of $X_\mathrm{max}$ and $R_\mu$ at $10$ EeV.}
\begingroup
\renewcommand{\arraystretch}{1} 
\begin{tabular}{l|cccc|}
\cline{2-5}
    & $\log_{10}(E_\mathrm{min}/\mathrm{eV})$ & $\log_{10}(E_\mathrm{max}/\mathrm{eV})$ & {$n$} & {$f_\mathrm{thres}$} \\ \hline
\multicolumn{1}{|c|}{\textsc{QGSJetII-04}}                  & $15$                                                           & $17$                                                           & 2                        & 1.0                                    \\ \hline
\multicolumn{1}{|c|}{\multirow{2}{*}{\textsc{EPOS-LHC}}}    & $15$                                                           & $18$                                                           & 1                        & 1.0                                    \\
\multicolumn{1}{|l|}{}                                      & $15$                                                           & $17$                                                           & 4                        & 1.0                                    \\ \hline
\multicolumn{1}{|c|}{\multirow{2}{*}{\textsc{Sibyll-2.3d}}} & $15$                                                           & $18$                                                           & 1                        & 1.0                                    \\
\multicolumn{1}{|l|}{}                                      & $15$                                                           & $17$                                                           & 4                        & 1.0                                    \\ \hline
\end{tabular}
\endgroup
\end{table}



\clearpage
\section{A {`Strangeball'} Extension to the Heitler-Matthews Model}
\label{sec:HeitlerMatthews}

From section~\ref{sec:impactEAS} we know that only a reduced fireball model --- { where no plasma is formed and the multiplicity is unaltered} 
--- could potentially resolve the muon puzzle. {Without the formation of a plasma state, it may be misleading to continue calling this a ``fireball''. Therefore, lacking a better word, we will refer to this reduced fireball as a ``strangeball''. It is interesting to note that the rope hadronization model for string fragmentation \cite{1984Biro, 2014Bierlich, 2022BierlichJetMod, 2022BierlichStrange} actually provided the best description of the ALICE strangeness enhancements in proton-proton collisions \cite{2017ALICEStrangenesspp} without invoking the formation of a plasma. Perhaps this could provide an alternative microscopic picture for strangeball formation.} 
For the remainder of this work we thus consider {a \emph{strangeball} model, which consists solely} of an appropriate swapping of pions and kaons, and thus only 
affects the muon number. An extension of the previous analysis to other CR energies could then be attempted with an analytic approach, starting from the Heitler-Matthews model \cite{\HeitlerMatthews}. 

\subsection{Derivation}

In the Heitler-Matthews model a hadronic air shower is modeled as consisting of charged and neutral pions. Neutral pions promptly decay to two photons, leaking energy from the hadronic component to EM showers. The charged pions produce further sets of charged and neutral pions until their energies fall below the critical energy --- defined as where the decay length becomes shorter than the interaction length --- at which point they decay to muons in a 1:1 ratio.

The number of muons thus corresponds to the number of charged pions at the end of a shower. Assuming the production of $n_\mathrm{mult}$ pions in an interaction, of which a fraction $r$ is charged, one has $(rn_\mathrm{mult})^k$ charged pions after $k$ interactions (or generations). With the further strong assumption of dividing the projectile energy equally over all produced pions, the critical generation $k_c$ (where charged pions reach the critical energy $E_c$) follows from $E_0/n_\mathrm{mult}^{k_c} = E_c$, with $E_0$ the energy of the CR primary. Combining these, the muon number at the end of the shower is given by
\begin{align}
    N_\mu = (r n_\mathrm{mult})^{k_c} = \left( \frac{E_0}{E_c} \right)^\beta \;, \label{eq:NmuHM}
\end{align}
where $\beta \equiv \log (rn_\mathrm{mult})/\log(n_\mathrm{mult})$ \cite{2010Ulrich}.

Swapping pions and kaons mainly induces a suppression of the produced number of neutral pions. Therefore, {strangeball} interactions differ from the Standard Model ones by an increased fraction of energy remaining in the hadronic component after an interaction. In the Heitler-Matthews model this fraction $r \equiv E_\mathrm{had}/E_\mathrm{proj}$ is constant, but in our extension we allow it to vary throughout the shower, taking at projectile energy $E$ the effective value
\begin{align}
    r_\mathrm{eff}(E) \equiv [1-p(E)]r_\mathrm{SM} + p(E) r_\mathrm{sb}  \, , \label{eq:refffb}
\end{align}
where $r_\mathrm{SM}$ and $r_\mathrm{sb}$ are the values for Standard Model and {strangeball} interactions, respectively, and $p(E)$ is the {strangeball}-production probability (eq.~(\ref{eq:probfireball})).

From $d \log(N_\mathrm{had}/N_\mathrm{tot})/dk$~$=$~$d \log((rn_\mathrm{mult})^k / n_\mathrm{mult}^k)/dk$ $= \log r$ we see that $r$ quantifies the generational change of the fraction of particles in the hadronic component. With an exponentiation and a multiplication by the total number of particles as fixed by the critical energy $N_\mathrm{tot} = E_0/E_c$, we can incorporate an energy-dependent $r$-value for the computation of the muon number:
\begin{align}
    N_\mu = \left(\frac{E_0}{E_c}\right) \exp \left[\int_{0}^{k_c} \log \{r_\mathrm{eff}(E)\} \; dk + \Delta_\mathrm{disc} \right] \;. \label{eq:muonderiv}
\end{align}
Here the upgrade of $k$ from a discrete to a continuous variable required the correction term $\Delta_\mathrm{disc} \equiv \tfrac{1}{2} \log(r_\mathrm{sb}/r_\mathrm{SM}) \, p(E_0)$, constituting an additional half generation of {strangeballs}. This takes into account that the type of interaction is determined by the incoming particles.

The multiplicity relates the generation $k$ to the energy $E$ of the particles at that generation. In the Heitler-Matthews model we have $E = E_0/n_\mathrm{mult}^k$, but at this point we can also consider an energy-dependent multiplicity. Taking a power-law,
\begin{align}
    n_\mathrm{mult}(E) = n_\mathrm{scale} \left(\frac{E}{1\; \mathrm{GeV}}\right)^b \;, \label{eq:PLmult}
\end{align}
we can solve the recurrence relation $E_{k+1} = E_k / n_\mathrm{mult}(E_k)$, which enables the computation of the Jacobian $dk/d\log E$ for the integral of eq.~(\ref{eq:muonderiv}).

Using our parametrization of the fireball {(i.e. strangeball)} production probability (eq.~(\ref{eq:probfireball})) along with eq.~(\ref{eq:refffb}) and the power-law multiplicity (eq.~(\ref{eq:PLmult})), we compute the muon number through eq.~(\ref{eq:muonderiv}) as
\begin{align}
    N_\mu = 
    \left(\frac{E_0}{E_c}\right) \left[ \frac{ x_c }{ x_0 } \right]^{c_1} \times
    \begin{cases}
    1 \vphantom{\mathlarger{\left( \frac{E_0}{E_\mathrm{min}} \right)^{\delta'(E_0)} }},  &\mathrm{if} \; E_0 \leq E_\mathrm{min},\\
    \mathlarger{\left( \frac{E_0}{E_\mathrm{min}} \right)^{\delta'(E_0)} }, &\mathrm{if} \; E_\mathrm{min} \leq E_0 \leq E_\mathrm{max},\\
    \mathlarger{\left( \frac{E_\mathrm{max}}{E_\mathrm{min}} \right)^{\delta'(E_\mathrm{max})} \left[ \frac{x_\mathrm{max}}{x_0} \right]^{c_2} }, \qquad \qquad &\mathrm{if} \; E_0 \geq E_\mathrm{max},
    \end{cases} \label{eq:NmufbintPLmultcomplete}
\end{align}
where we defined 
\begin{align}
    c_1 \equiv \frac{\log r_\mathrm{SM}}{\log(1-b)}, \qquad
    c_2 \equiv \frac{\log(r_\mathrm{sb}/r_\mathrm{SM})}{\log(1-b)}, \label{eq:slopes}
\end{align}
\vspace{-.5cm}
\begin{align}
    x_i \equiv \log \left( n_\mathrm{scale} \left(\frac{E_i}{1 \; \mathrm{GeV}}\right)^b \right), \qquad i \in \{c, 0, \mathrm{min}, \mathrm{max}\}, \label{eq:logterms}
\end{align}
\vspace{-.4cm}
\begin{align}
    \delta'(E) \equiv - \frac{p(E)}{n+1} \frac{c_2}{x_\mathrm{min}/b} \,\, {}_2 F_1 \left(1, 1+n; 2+n; \frac{-\log(E/E_\mathrm{min})}{x_\mathrm{min}/b}\right) + \frac{1}{2} \frac{\log(r_\mathrm{sb}/r_\mathrm{SM})}{\log(E_\mathrm{max}/E_\mathrm{min})} p(E)^\mathlarger{\frac{n-1}{n}}, \label{eq:edepslope}
\end{align}
for compactness and readability. The function $_2 F_1 (a, b; c; x)$ is the hypergeometric function.

The three energy regimes of $p(E)$ is reflected here as a collection of distorted power-laws. Below $E_\mathrm{min}$ the power-law of the Heitler-Matthews model (eq.~(\ref{eq:NmuHM})) transforms to a power-law of logarithmic terms (eq.~(\ref{eq:logterms})) due to the energy-dependent multiplicity. The transition region from no {strangeballs} to only {strangeballs} is characterized by a further power-law with an energy-dependent slope (eq.~(\ref{eq:edepslope})). Above $E_\mathrm{max}$ one again obtains the transformed power-law with logarithmic terms, but then with the slope adjusted for {strangeballs} (eq.~(\ref{eq:slopes})).

\subsection{Parameter Estimation}

Summarizing the parameters, we have the {strangeball} settings $\{E_\mathrm{min}, E_\mathrm{max}, n \}$ (also through $p(E)$ in eq.~(\ref{eq:edepslope})), the physical quantities $\{r_\mathrm{SM}, r_\mathrm{sb}, n_\mathrm{scale}, b, E_c\}$, and the energy $E_0$ of the primary CR. For the application of this framework to Auger data we need to estimate the physical quantities.

Since the hadronic energy fraction and the multiplicity are properties of individual interactions, we attempted to obtain $\{r_\mathrm{SM}, r_\mathrm{sb}, n_\mathrm{scale}, b \}$ from the high-energy hadronic interaction models directly. We used the CRMC software package \cite{CRMC} as a uniform interface to \textsc{QGSJetII-04}, \textsc{EPOS-LHC}, and \textsc{Sibyll-2.3c}\footnote{At the time of this study, the updated version \textsc{Sibyll-2.3d} was not yet available in CRMC.} \cite{2019Sibyll23c_PRD}, simulating $10^4$ fixed-target collisions of energetic proton and $\pi^+$ projectiles with stationary proton and nitrogen targets, varying the projectile energy from $10^2$ GeV to $10^{11}$ GeV in steps of factors of 10. 

From each collision we computed the fraction of the projectile energy that is carried away by hadronic secondaries, counting all particles but $\{ \pi^0, e^\pm, \gamma \}$ for the Standard Model, and all particles but $\{ K^0_{L/S}, e^\pm, \gamma \}$ for the {strangeball} model. Changing the projectile energy or the type of interacting particles induce variations of $\langle r_\mathrm{SM} \rangle$ ($\langle r_\mathrm{sb} \rangle$) within 0.1 (0.05). In our simplified approach we ignore these variations and computed the global averages, giving equal weights to all energies and projectile-target combinations, resulting in the parameters listed in table~\ref{tab:paramestimates}. We further computed the total multiplicity of each collision, finding the averages among many collisions to follow a power-law with projectile energy, for each projectile-target combination. Averaging these power laws with equal weights (point-wise at each energy) resulted in a global power law, of which the parameters are also listed in table~\ref{tab:paramestimates}.

\newcommand*{\UniWidth}[1]{\parbox[c]{1cm}{#1}}%
\begin{table}[h]
\begingroup
\centering
\caption{\label{tab:paramestimates} Estimates of the parameters related to physical quantities for the evaluation of eq.~(\ref{eq:NmufbintPLmultcomplete}), using both \textsc{CRMC} and \textsc{Conex} simulations and various hadronic interaction models.}
\medskip

\renewcommand{\arraystretch}{1.3} 

\makebox[\linewidth][c]{ 
\begin{tabular}{cc|ccccc|}
\cline{3-7}
& & \UniWidth{$r_\mathrm{SM}$} & \UniWidth{$r_\mathrm{sb}$} & \UniWidth{$n_\mathrm{scale}$} & {$b$} & $E_c$ [GeV] \\ \hline
\multicolumn{1}{|c|}{\multirow{3}{*}{\rotatebox[origin=c]{90}{\textsc{CRMC}}}}      & \textsc{QGSJetII-04} &    0.781             & 0.937                &  5.69                  & 0.193                      & -               \\
\multicolumn{1}{|c|}{}  & EPOS-LHC              & 0.788     & 0.930     & 7.70      & 0.166                      & -               \\
\multicolumn{1}{|c|}{}  & \textsc{Sibyll-2.3c}  & 0.803     & 0.921     & 6.74      & 0.173         & -               \\ \hline
\multicolumn{1}{|c|}{\multirow{3}{*}{\rotatebox[origin=c]{90}{\textsc{{Conex}}}}} & \textsc{QGSJetII-04} & 0.509          & 0.720          & 968              & $8.68 \cdot 10^{-2}$ & 136           \\
\multicolumn{1}{|c|}{}                                    & EPOS-LHC    & 0.550          & 0.764          & 3820               & $2.58 \cdot 10^{-3}$ & 154           \\
\multicolumn{1}{|c|}{}                                    & \textsc{Sibyll-2.3d} & 0.565          & 0.736          & 3230               & $3.92 \cdot 10^{-5}$ & 151           \\ \hline
\end{tabular}
}
\endgroup
\end{table}
\vspace{0.3cm}

\begin{figure}
\centering
    \includegraphics[width=1\textwidth]{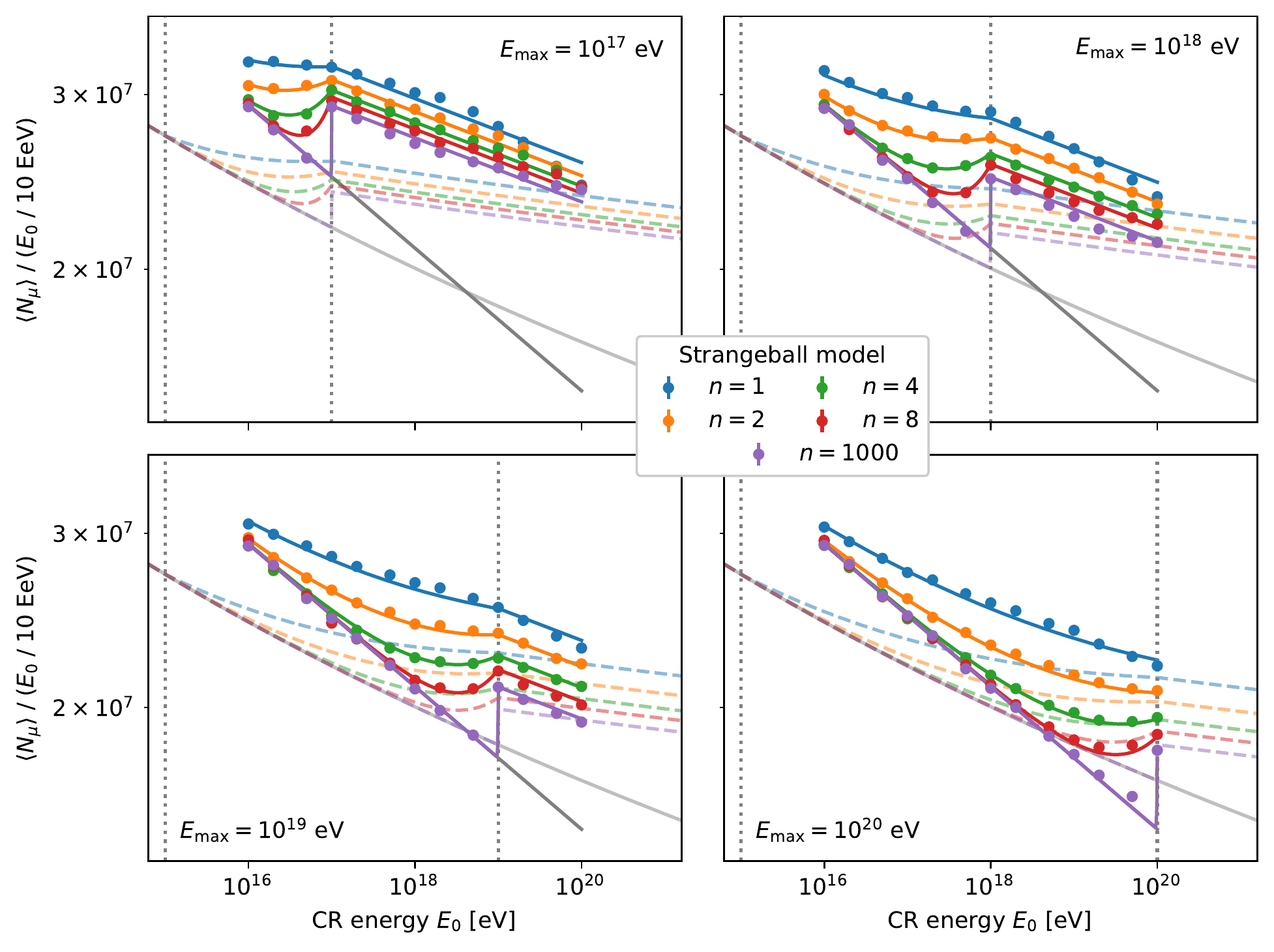}
    \caption{\label{fig:Conexfit} Energy dependence of the average muon number for various {strangeball} settings with \textsc{EPOS-LHC} as obtained with \textsc{Conex} simulations (data points) fitted with eq.~(\ref{eq:NmufbintPLmultcomplete}) (solid lines), and using \textsc{CRMC}-inferred parameters directly with eq.~(\ref{eq:NmufbintPLmultcomplete}) (dashed lines). The gray lines correspond to the no {strangeball} case, and we fixed $E_\mathrm{min}=10^{15}$ eV.}
\end{figure}

Since we implicitly consider the hadronic shower component to consist of more than just charged pions, the definition of a single critical energy becomes ambiguous. Therefore, instead of estimating its value from first principles (through the interaction and decay lengths), we treat it as a normalization parameter. Setting it to $E_c = 220$ GeV and using the remaining parameters from \textsc{CRMC}, the number of muons as a function of primary energy $E_0$ from eq.~(\ref{eq:NmufbintPLmultcomplete}) is given by the dashed lines of figure~\ref{fig:Conexfit} for \textsc{EPOS-LHC} and $E_\mathrm{min}=10^{15}$ eV.

We ran an additional set of \textsc{Conex} simulations to evaluate the validity of this approach. Only considering proton CRs, we explored the {strangeball} phase-space varying $E_\mathrm{min} \in \{10^{14}, 10^{15}, 10^{16}\}$ eV, $E_\mathrm{max} \in \{10^{17}, 10^{18}, 10^{19}, 10^{20}\}$ eV, and $n \in \{1, 2, 4, 8, 1000\}$. The energy-dependence of the average muon number is sampled with $E_0 \in \{ 1, 2, 5, 10, 20, 50, 100,$ $200, 500, 1000, 2000, 5000, 10000 \}~\times 10^{16}$ eV, and the simulations are repeated for the hadronic interaction models \textsc{QGSJetII-04}, \textsc{EPOS-LHC}, and \textsc{Sibyll-2.3d}{, simulating 2100 showers for each setting}. The resulting values for \textsc{EPOS-LHC} and $E_\mathrm{min}=10^{15}$ eV are indicated by the { points (with negigible statistical uncertainty)} in figure~\ref{fig:Conexfit}.

Independent of our choice of $E_c$, the muon numbers from \textsc{Conex} simulations are not reproduced by eq.~(\ref{eq:NmufbintPLmultcomplete}) when using the parameters inferred from CRMC simulations. This implies that our attempt to connect microscopic parameters to a macroscopic observable with an analytic model is too simplistic. This inadequacy can most likely be traced back to the unphysical Heitler-Matthews assumption of equally dividing the projectile energy over all secondaries, which is in direct contradiction with the CRMC spectra. One may consider combining our {strangeball} extension with the extension presented in \cite{2017GrimmProceedings} to take into account leading particle effects.


Instead, since the functional form given by the \textsc{Conex} simulations seems to be reproduced, we fitted eq.~(\ref{eq:NmufbintPLmultcomplete}) directly to the results of these simulations (fixing $E_\mathrm{min}=10^{15}$ eV; 260 data points), as shown by the solid lines in figure~\ref{fig:Conexfit}. This gave a surprisingly good fit, where we obtained a single set of parameters for each model as listed in table~\ref{tab:paramestimates}. It thus seems that realistic effects such as that of leading particles can be absorbed into these parameters, making them effective and their physical interpretation should be met with some caution. The same parameters provided a similarly good description of the \textsc{Conex} simulations with $E_\mathrm{min} \in \{10^{14}, 10^{16} \}$ eV, implying that eq.~(\ref{eq:NmufbintPLmultcomplete}) can be used to interpolate the average muon number $N_\mu$ in the primary energy $E_0$, as well as the three {strangeball} settings $E_\mathrm{min}$, $E_\mathrm{max}$, and $n$.

\subsection{Mass Dependence}

\begin{figure}[h]
    \centering
    \includegraphics[width=0.9\textwidth]{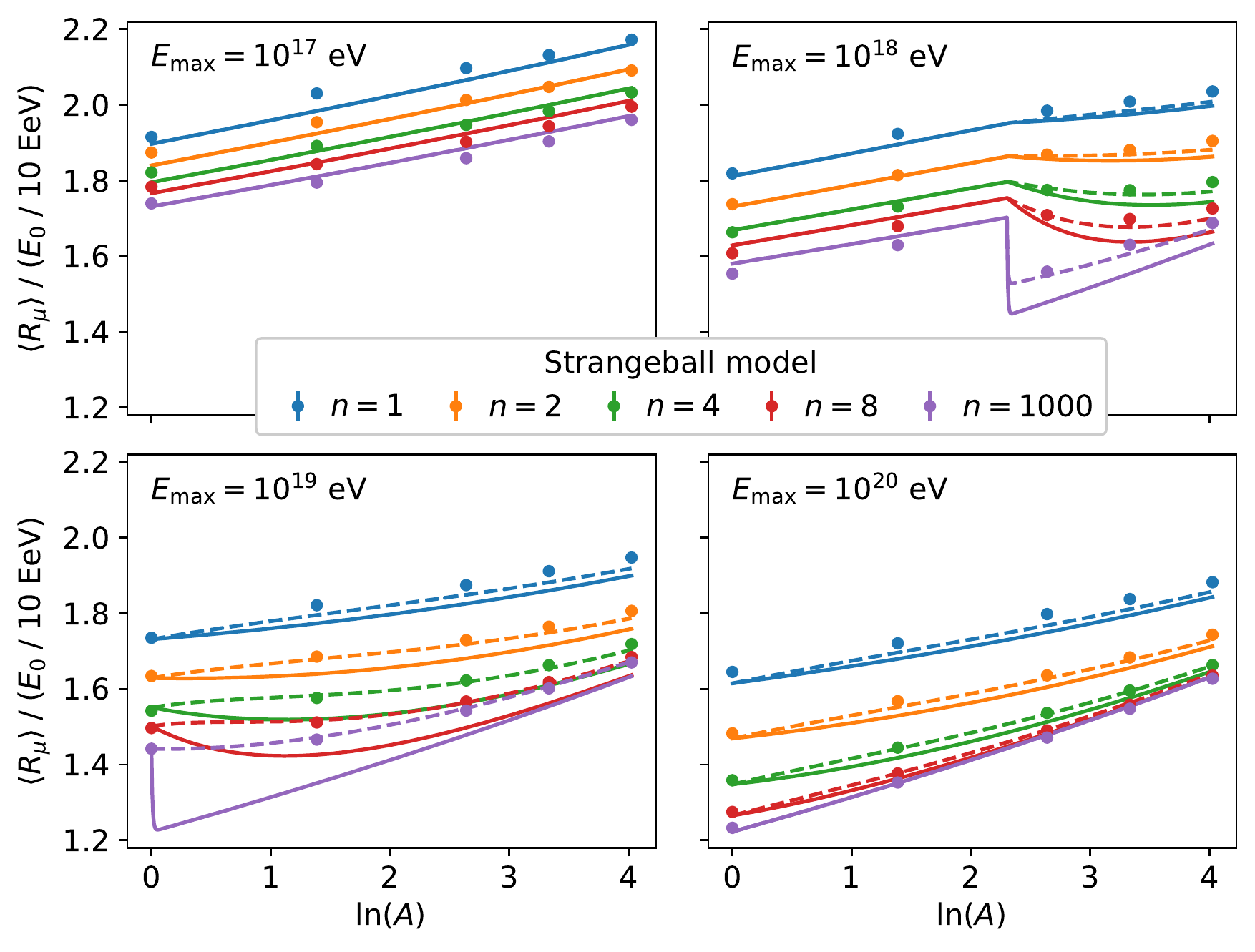}
    \caption{\label{fig:superpos} Mass dependence of the average muon number at 10 EeV for various {strangeball} settings obtained from \textsc{EPOS-LHC} \textsc{Conex} simulations (data points) and compared with our analytic model (eq.~(\ref{eq:NmufbintPLmultcomplete}) with \textsc{Conex} parameters of table~\ref{tab:paramestimates}) by applying the superposition principle (eq.~(\ref{eq:superpos}), solid lines) and when including our correction factor (with eq.~(\ref{eq:superposadj}), dashed lines).}
\end{figure}

Now we need to introduce a mass dependence to our analytic model by way 
of the superposition principle:
\begin{align}
    N_\mu(E_0,A) = A \cdot N_\mu(E_0/A, 1) \;. \label{eq:superpos}
\end{align}
This states that the muon number from a shower initiated by a CR with energy $E_0$ and mass $A$ corresponds to that of $A$ protons, each a factor $A$ lower in energy. We assume that the total energy, rather than that per nucleon, is decisive for the production of a {strangeball} state. Note that in this case the conventional superposition principle underestimates the {strangeball} production from nuclei. This becomes especially apparent when considering a rapidly changing {strangeball} production probability $p(E)$, as illustrated in figure~\ref{fig:superpos} (e.g., the line with $E_\mathrm{max}=10^{19}$ eV and $n=1$).

We introduce a correction factor to the right-hand side of eq.~(\ref{eq:superpos}),
\begin{align}
    u + (1-u) \cdot \left[r_\mathrm{sb}/r_\mathrm{SM}\right]^{\Delta p / 2} \;, \label{eq:superposadj}
\end{align}
where $u$ denotes the fraction of the nucleus that does not participate in the first inelastic interaction (`spectator nucleons'), and is thus unaffected by the {strangeball} state. The remainder of the nucleus finds its muon production enhanced by half a generation of additional {strangeballs}, weighted by the difference in {strangeball}-production probability for the nucleus and its nucleons: $\Delta p \equiv p(E_0) - p(E_0/A)$. This correction factor is equivalent to revising $\Delta_\mathrm{disc}$ of eq.~(\ref{eq:muonderiv}) to be evaluated at the energy of the nucleus rather than that of its nucleons.

We found a good agreement with \textsc{EPOS-LHC} \textsc{Conex} simulations when setting $u=1-1/\sqrt{A}$, as indicated by the dashed lines in figure~\ref{fig:superpos}. This corresponds to approximately $\sqrt{A}$ nucleons interacting inelastically. While a slight deviation remains for the other two models, we expect our approximation to suffice for the current study.

\section{Application to Auger Data}
\label{sec:ApplyAuger}

Equipped with eq.~(\ref{eq:NmufbintPLmultcomplete}), its parameters, and a mass dependence, we looked for {strangeball} settings that reproduce Auger data. First, we verified with \textsc{Conex} simulations that the statistical moments of $X_\mathrm{max}$ are unaffected by the {strangeball} model. We simulated proton and iron showers with energy in the range from $10^{17}-10^{20}$ eV and {strangeball} settings $E_\mathrm{min}=10^{15}$ eV, $E_\mathrm{max} \in \{10^{17}, 10^{18}, 10^{19}, 10^{20}\}$ eV, and $n\in \{ 1, 2, 4, 8, 1000\}$, for each hadronic interaction model. We did not find a significant deviation with respect to simulations without {strangeballs}, justifying our approach to only consider changes to the muon number. This puts us in a situation where we can assume the composition from $X_\mathrm{max}$ to be the true composition 
and that we only need to adjust the muon predictions until they 
give the same picture. 

\begin{figure}[h]
    \vspace{0.3cm}
    \centering
    \includegraphics[width=0.55\textwidth]{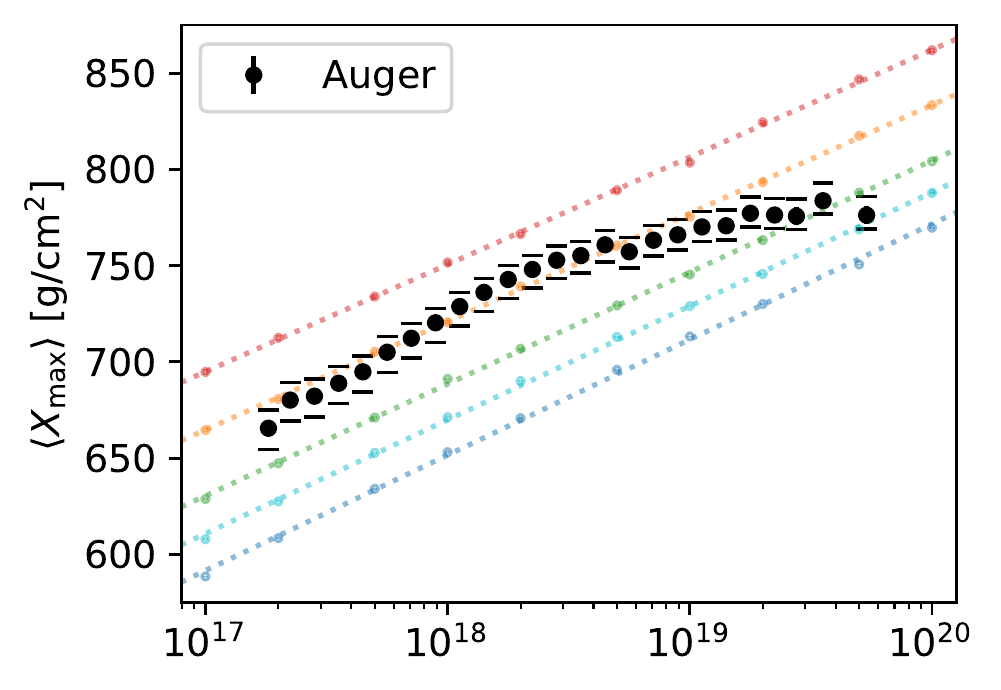}
    \includegraphics[width=1\textwidth]{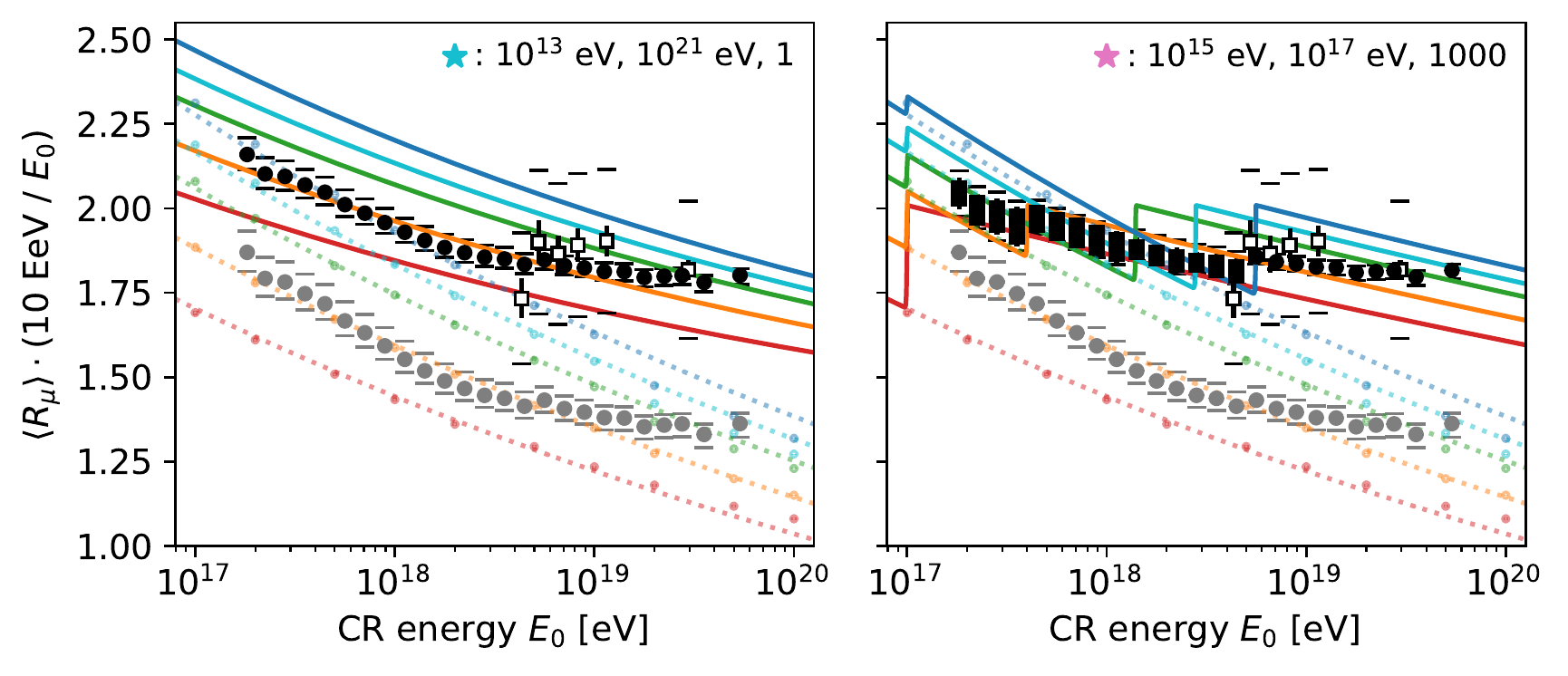}
    \caption{\label{fig:interpAuger3plot} Composition inference from Auger data (error bars) on $\langle X_\mathrm{max} \rangle$ ({top}) and $\langle R_\mu \rangle$ ({bottom}) using \textsc{EPOS-LHC} and the {strangeball} model. The {strangeball} model (solid lines) leaves $\langle X_\mathrm{max} \rangle$ unaffected, data on which (black error bars) can thus be interpreted within the Standard Model (dotted lines). In the {bottom} plots a direct comparison with $\langle R_\mu \rangle$ data ({white square} error bars) follows from mapping $\langle X_\mathrm{max} \rangle$ data to $\langle R_\mu \rangle$ within the Standard Model ({gray} error bars) and {the following} two {strangeball} scenarios ({black} error bars): $E_\mathrm{min}=10^{13}$ eV, $E_\mathrm{max}=10^{21}$ eV and $n=1$ ({bottom left}), and $E_\mathrm{min}=10^{15}$ eV, $E_\mathrm{max}=10^{17}$ eV and $n=1000$ ({bottom right}). The line colors correspond to various nuclei: proton (red), helium (orange), nitrogen (green), silicon (cyan), and iron (blue).}
\end{figure}

By mapping data on $\langle X_\mathrm{max} \rangle$ to $\langle R_\mu \rangle$ we can directly quantify the size of the muon discrepancy, as visualized in figure~\ref{fig:interpAuger3plot} for \textsc{EPOS-LHC}. In the {top plot} we find the $\langle X_\mathrm{max} \rangle$ predictions to follow power-laws in energy, enabling a straightforward interpolation to the energies of the data points. The interpolation in energy of $\langle R_\mu \rangle$ comes from our analytic model, with which we mapped the $\langle X_\mathrm{max} \rangle$ data to the {bottom plots} following the same procedure as in section~\ref{sec:impactEAS}. These plots correspond to opposite extremes of the {strangeball} settings: a gradual introduction of {strangeballs} starting at low energies ($10^{13}$ eV; left), or an abrupt introduction around $10^{17}$ eV (right). Considering only these data, both scenarios seem to resolve the muon puzzle.

A distinction could be made by including information on the relative muon fluctuations (e.g., \cite{2021AugerMuonFluc}), which reflects the physics of the first interaction. In appendix~\ref{app:muonfluc} we explored the effect of {strangeballs} on the energy dependence of these fluctuations following a modeling similar to that of section~\ref{sec:HeitlerMatthews}. We found that {strangeballs} in the first interaction tend to decrease the relative muon fluctuations, which could lead to shifting the muon puzzle to these fluctuations. 
This effectively puts a lower limit on $E_\mathrm{max}$, requiring a 100\% probability of producing {strangeballs} to be only allowed at energies above what is measured by the Pierre Auger Observatory.

\begin{figure}[t]
    \centering
    \includegraphics[width=1\textwidth]{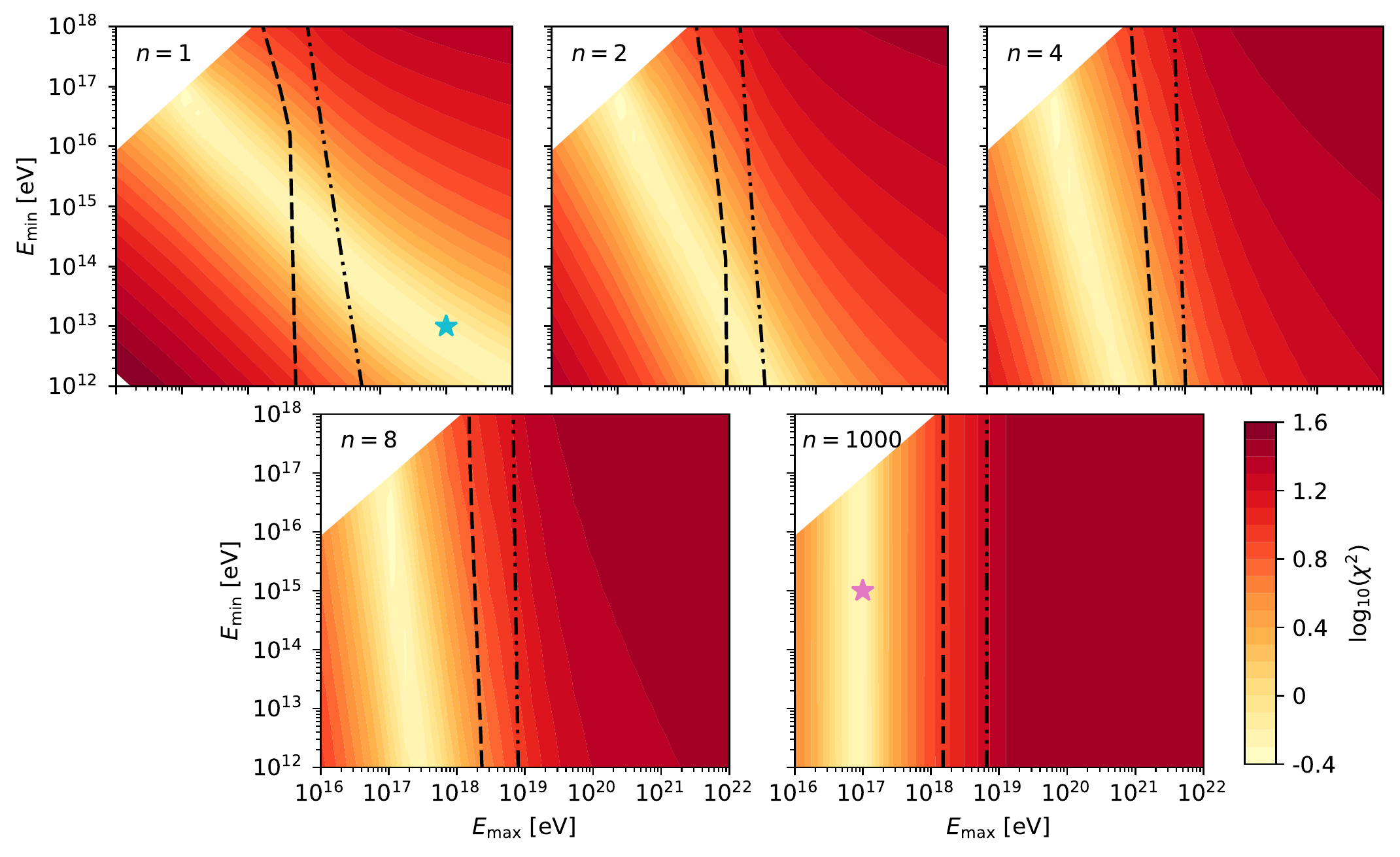}
    \caption{\label{fig:interpAugerSCAN} {Strangeball} parameter-space exploration of the compatibility of the composition inference of Auger data on $\langle X_\mathrm{max} \rangle$ and $\langle R_\mu \rangle$ as quantified by the test statistic $\chi^2$ (eq.~(\ref{eq:chisqconsist})) for \textsc{EPOS-LHC}. A lower $\chi^2$ implies a better compatibility. The inset stars correspond to the central (cyan) and right (purple) plots in figure~\ref{fig:interpAuger3plot}. The black lines are lower limits on $E_\mathrm{max}$ required by data on the muon fluctuations.}
\end{figure}

For a complete exploration of the phase-space of {strangeball} settings 
we fixed $n \in \{1, 2, 4, 8,$ $1000 \}$ and varied both $10^{12} \leq E_\mathrm{min}/\mathrm{eV} \leq 10^{18}$ and $10^{16} \leq E_\mathrm{max}/\mathrm{eV} \leq 10^{22}$. As visualized in figure~\ref{fig:interpAugerSCAN} for \textsc{EPOS-LHC} (similar figures were obtained for \textsc{QGSJetII-04} and \textsc{Sibyll-2.3d}) we computed for each setting a chi-squared statistic to quantify the agreement between data on $\langle X_\mathrm{max} \rangle$ and $\langle R_\mu \rangle$:
\begin{align}
        \chi^2(E_\mathrm{min}, E_\mathrm{max}, n) \equiv
        \sum_{i=1}^{6} \frac{ \left[ \langle R_\mu \rangle_{\langle X_\mathrm{max} \rangle_\mathrm{data}}(E_i; E_\mathrm{min}, E_\mathrm{max}, n) - \langle R_\mu \rangle_{\mathrm{data},i} \right]^2}{\delta\langle R_\mu \rangle_{\mathrm{syst},i}^2 + \delta \langle R_\mu \rangle_{\mathrm{stat},i}^2} \;.
    \label{eq:chisqconsist}
\end{align}
The sum runs over the six muon data points and $\langle R_\mu \rangle_{\langle X_\mathrm{max} \rangle_\mathrm{data}}(E_i; E_\mathrm{min}, E_\mathrm{max}, n)$ represents the mapped data on $\langle X_\mathrm{max} \rangle$, which is subsequently linearly interpolated in $\log E$ to the energies $E_i$ of the data points $\langle R_\mu \rangle_{\mathrm{data},i}$. We neglect the uncertainty on $\langle X_\mathrm{max} \rangle$ data and consider the total uncertainty on $\langle R_\mu \rangle$ data to correspond to a quadratic sum of the systematic $\delta \langle R_\mu \rangle_{\mathrm{syst},i}$ and statistical $\delta \langle R_\mu \rangle_{\mathrm{stat},i}$ uncertainties. Note that a division of $\langle R_\mu \rangle$ by energy is canceled by the uncertainty term.

A resolution of our muon puzzle follows from enhancing the $\langle R_\mu \rangle$ predictions to a certain plateau, which only corresponds to a single constraint on the {strangeball} settings. This is reflected by the valleys of solutions in figure~\ref{fig:interpAugerSCAN} for each value of $n$. Lower limits on $E_\mathrm{max}$ from data on the relative muon fluctuations are indicated by the black lines, to the left of which proton predictions fall more than $1~\sigma$ below these data (and the expected heavier composition corresponds to an even larger tension). Combining these constraints we find that with fluctuations limiting the number of {strangeballs} in the first interaction, {strangeballs} need to be present at lower energies, to reach the plateau and resolve the muon puzzle. This roughly excludes scenarios with $n>1$, favoring the gradual (blue star) over the abrupt (purple star) solution.

\begin{figure}[h]
    \centering
    \makebox[1\textwidth]{
        \includegraphics[width=1.1\textwidth]{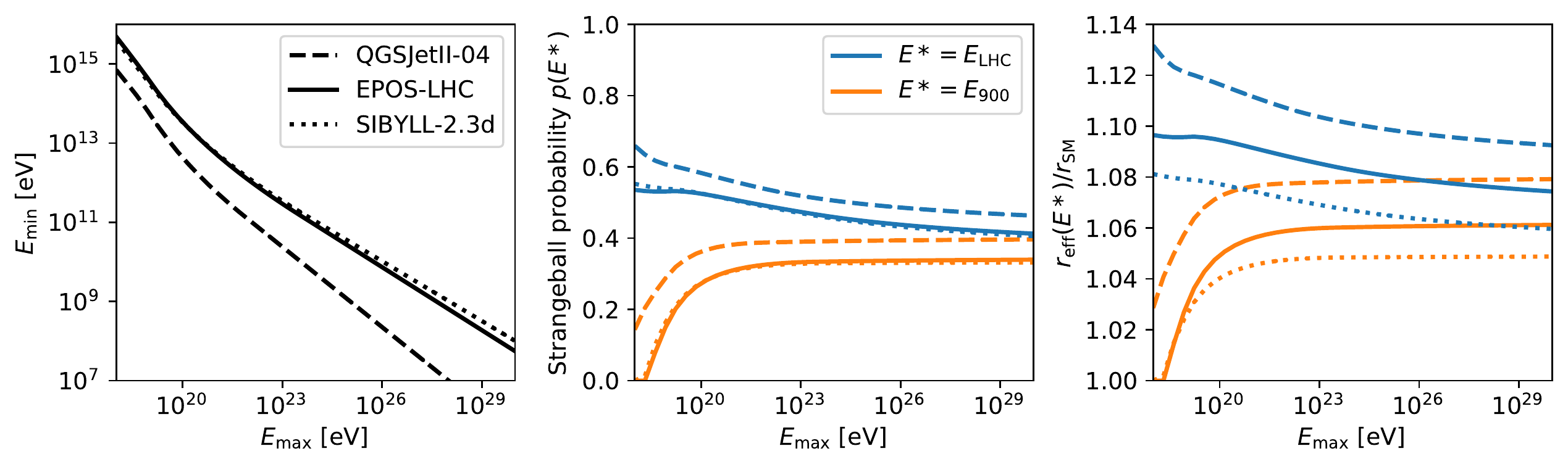}
    }
    \caption{\label{fig:interpAugerbestfit} \emph{Left:} {Strangeball} settings resolving the muon puzzle without violating constraints from the muon fluctuations (i.e., $E_\mathrm{max} \gtrsim 10^{19}$ eV). \emph{Center and right:} Conversion of these {strangeball} settings to the {strangeball}-initiation probability (center, eq.~(\ref{eq:probfireball})) and effective enhancement of the hadronic energy fraction (right, eq.~(\ref{eq:refffb})) at LHC (blue, $E_\mathrm{LHC} \approx 10^{17}$ eV) and Tevatron (orange, $E_\mathrm{Tev} \approx 10^{15}$ eV) energies.}
\end{figure}

Considering only $n=1$, the combinations of $E_\mathrm{min}$ and $E_\mathrm{max}$ that minimize $\chi^2$ are shown on the left of figure~\ref{fig:interpAugerbestfit} for each hadronic interaction model. Note that $E_\mathrm{max}$ starts at $10^{18}$ eV and thus already roughly takes into account the constraint from the fluctuations. The lower values for \textsc{QGSJetII-04} reflects its lower muon numbers. We converted these solutions to the probability of producing {strangeballs} (central plot) and the effective enhancement to the hadronic energy fraction (right plot) at LHC ($\sqrt{s_\mathrm{LHC}} = 13$ TeV $\Leftrightarrow E_\mathrm{LHC} = 8.45 \cdot 10^{16}$ eV) and Tevatron ($\sqrt{s_\mathrm{Tev}} = 2$ TeV $\Leftrightarrow E_\mathrm{Tev} = 2 \cdot 10^{15}$ eV) energies. Only for the lowest allowed values of $E_\mathrm{max}$ may {strangeball} effects be invisible at Tevatron energies. More compatible 
with the constraints from the fluctuations, we find any solution to the muon puzzle to require $35-40$\% ($40-45$\%) of the interactions to be {strangeballs} at Tevatron (LHC) energies, corresponding to an effective increase of the hadronic fraction of $5-8$\% ($6-9$\%).

Some care is needed in the interpretation of these results. Solutions to the muon puzzle in terms of $E_\mathrm{min}$ and $E_\mathrm{max}$ combinations and the subsequent conversion to a {strangeball} probability rely on the correct interpolation (and extrapolation) of our analytic model in terms of the CR energy and the {strangeball} settings. We verified this with \textsc{Conex} simulations for a subset of settings ($10^{14} \leq E_\mathrm{min}/\mathrm{eV} \leq 10^{16}$ and $10^{17} \leq E_\mathrm{max}/\mathrm{eV} \leq 10^{20}$), as in, e.g., figure~\ref{fig:Conexfit}. Many of the solutions in figure~\ref{fig:interpAugerbestfit} are outside this range, but, given the physical assumptions going into our model, we do not expect large uncertainties associated to these extrapolations. For the subsequent conversion to an effective increase of the hadronic energy fraction we employed eq.~(\ref{eq:refffb}) with the parameters listed in table~\ref{tab:paramestimates}. Since the parameters inferred from \textsc{Conex} simulations are likely unphysical, with in particular the very high multiplicities, we used the representative parameters { found with} \textsc{CRMC} simulations instead. 

\section{Implications for LHC measurements}
\label{sec:LHCimplications}


The hadronic energy fraction $r$ is not a directly measurable quantity at collider experiments. Therefore, it is worthwhile {to} translate our results to observables of {the} relevant detectors at the LHC. {For this we need to explicitly specify the model under consideration.}
The analytic model of section~\ref{sec:HeitlerMatthews} is agnostic towards the precise origin of enhancing $r${, which implies that t}he results in the right-most plot of figure~\ref{fig:interpAugerbestfit} could be regarded as independent of the model behind the enhancement (e.g., swapping). {Furthermore, a phenomenological model valid for air showers may not necessarily be valid for collider experiments. Nevertheless, for the sake of consistency and simplicity, we will stick to modeling strangeballs by swapping pions for kaons.}

We would further like to point out that our focus on $r$ represents a generic approach to resolving the muon puzzle. This follows from an assumed proportionality between the energy kept in the hadronic shower component and the produced number of muons. However, by swapping particles or otherwise adjusting the hadronic particle spectra one may also enhance the efficiency of converting hadronic energy into the production of muons. To analyze this effect on the muon number it is convenient to reverse the problem and study their genealogy (see, e.g., refs.~\cite{1997Hillas,2021Reininghaus}), but such considerations are beyond the scope of this paper. 

\begin{figure}[h]
    \includegraphics[width=1\textwidth]{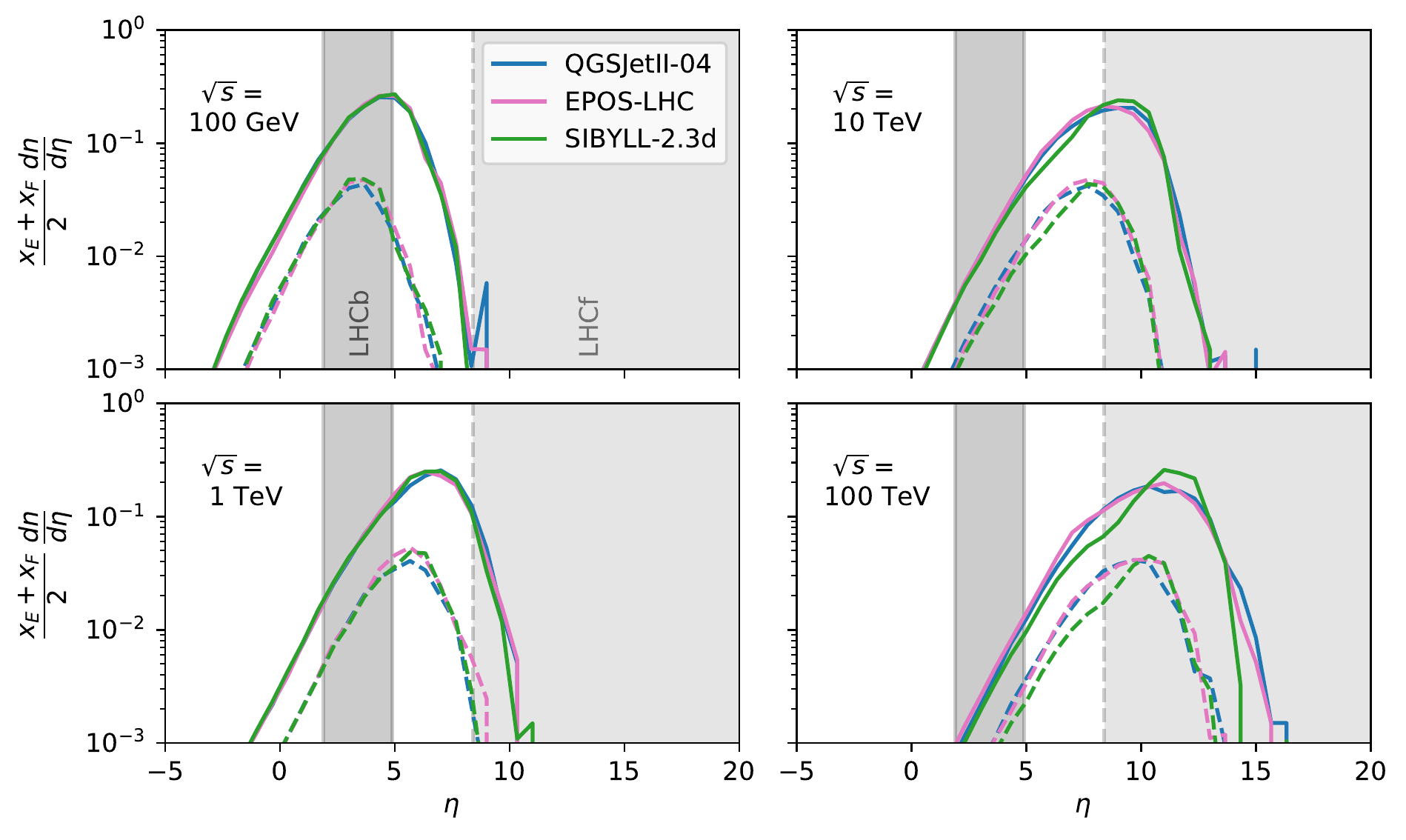}
    \caption{\label{fig:rapidity} Weighted {pseudo}rapidity distributions of all (solid lines) and EM-only (including $\pi^0$, dashed lines) secondaries from proton-proton collisions at various center-of-mass energies for the three hadronic interaction models{, computed with CRMC}. The weight corresponds to the fraction of projectile energy carried by secondaries when boosted to the fixed-target rest frame, {given by} eq.~(\ref{eq:Eweight}). The pseudorapidity acceptance regions of LHCb ($1.9 < \eta < 4.9$ \cite{2021MuonPuzzle}) and LHCf ($|\eta| > 8.4$ \cite{2016LHCf}) are indicated in gray.}
\end{figure}

The part of the air shower evolution relevant to the muon number are the secondaries that carry {away} most of the projectile energy $E_\mathrm{proj}$:
\begin{align}
    \frac{E'}{E_\mathrm{proj}} \approx \frac{E+p_z}{\sqrt{s}} = \frac{x_E + x_F}{2} \;. \label{eq:Eweight}
\end{align}
{Here $E$ and $p_z$ are respectively the energy and ($z$-)momentum of the secondary.} The prime denotes the Earth's rest frame, and the first equality follows from a boost to the center of mass frame (where $E'=\gamma(E+\beta p_z)$, with $\gamma=\sqrt{s}/(2m_p)$ and $\beta \approx 1$, and $E_\mathrm{proj} \approx s/(2m_p)$). 
The second step follows from the definitions of $x_E \equiv 2E/\sqrt{s}$ and Feynman-$x$ $x_F \equiv 2p_z/\sqrt{s}$. Since collider experiments detect secondaries in specific pseudorapidity intervals (and above energy thresholds), it is instructive to identify which pseudorapidity intervals contain most of the projectile energy. We show in figure~\ref{fig:rapidity} the pseudorapidity distributions computed with CRMC of all (solid lines) and only EM secondaries (dashed lines) from proton-proton interactions for various CM energies, weighted by eq.~(\ref{eq:Eweight}). 
There is a clear positive correlation between the CM energy and the pseudorapidity at which the distribution peaks, implying that different forward regions are relevant at different energies. An enhancement of $r$ corresponds to reducing the area underneath the EM spectrum (dashed lines). In principle, pseudorapidity regions away from the peaks need not be affected, but this requires fine-tuning that may be difficult to reconcile with an underlying physical model. Therefore, we consider the swapping of pions and kaons to be equivalently present in all kinematic regions. Note that since new physics would primarily arise in the constrained central regions, our approach could be regarded as conservative.

In the following we investigate the effect of swapping pions and kaons on the predictions of cosmic-ray hadronic interaction models in phase-space regions relevant for the LHCf and LHCb detectors. In particular, we assess whether current measurements permit an $O(40\%)$ of swapping as we found to be required for solving the muon puzzle (see the central plot in figure~\ref{fig:interpAugerbestfit}).


\subsection{LHCf}

\begin{figure}[h]
    \centering
    \makebox[1\textwidth]{
        \includegraphics[width=1.1\textwidth]{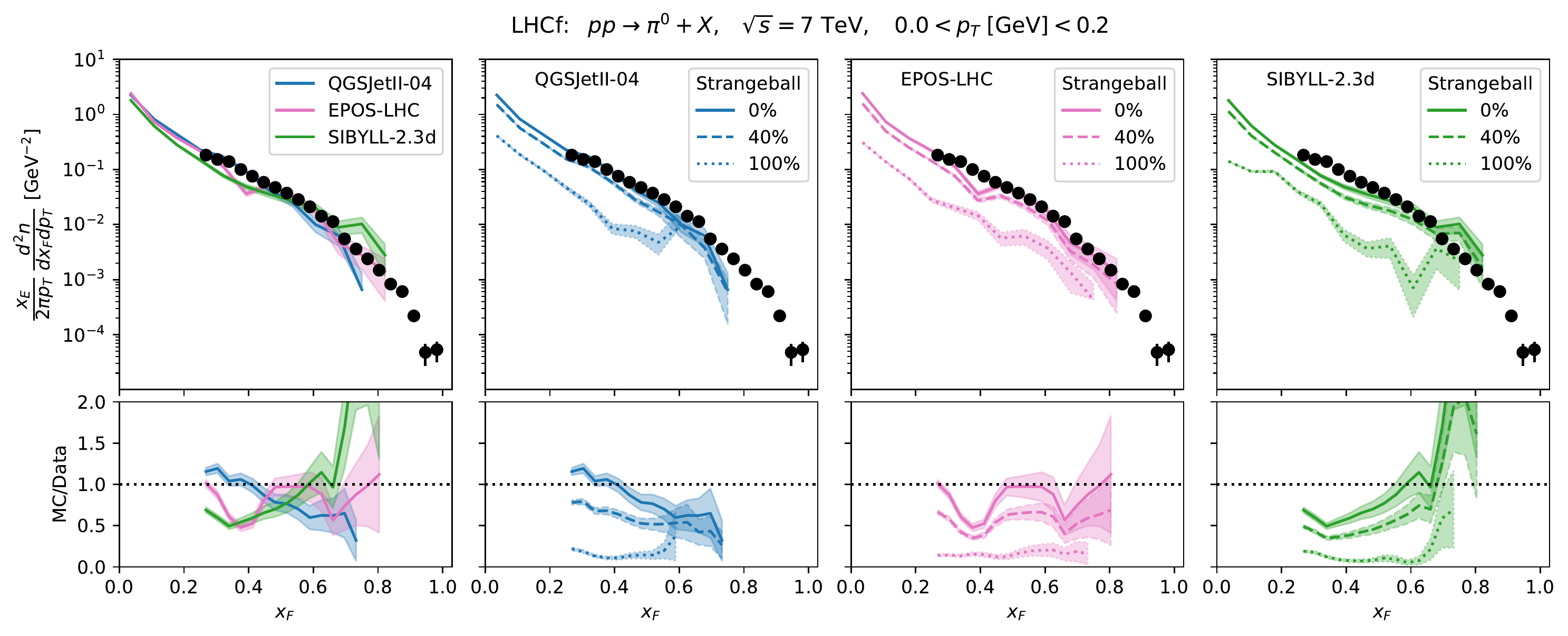}
    }
    \caption{\label{fig:LHCf} The neutral pion yield from proton-proton collisions at $\sqrt{s}$~=~7~TeV as measured by the LHCf detector \cite{2016LHCf} (data points) and retrieved from various models {with CRMC} (lines), for the $p_T$-range of 0.0 to 0.2 GeV. For clarity, we separately visualize the effect of 40\% and 100\% {strangeballs} for each of the hadronic interaction models. The bands correspond to $1\,\sigma$ Monte Carlo uncertainties.}
\end{figure}

The LHCf measurements of the neutral pion yield from $\sqrt{s} = 7$ TeV proton-proton collisions \cite{2016LHCf} constrain the energy going into the EM component, i.e. the complement of $r$. 
A comparison of these measurements with model predictions is shown in figure~\ref{fig:LHCf}, as well as the effect of a 40\% and 100\% {strangeball} (i.e., swapping pions and kaons). Note that here a 100\% {strangeball} simply corresponds to the neutral kaon spectrum (both $K_S^0$ and $K_L^0$). 
At this energy the LHCf detector is only sensitive to the high-rapidity tail of the EM distribution (see figure~\ref{fig:rapidity}), making only the $x_F < 0.4$ region relevant in practice for the muon number.
While a 40\% {strangeball} induces a significant suppression to the spectra, these deviations appear to be of the same order of magnitude as the model differences. Therefore, with the current theoretical uncertainties one cannot exclude a 40\% {strangeball} on the basis of these data.

\subsection{LHCb}

\begin{figure}[h]
    \centering
    \makebox[1\textwidth]{
        \includegraphics[width=1.1\textwidth]{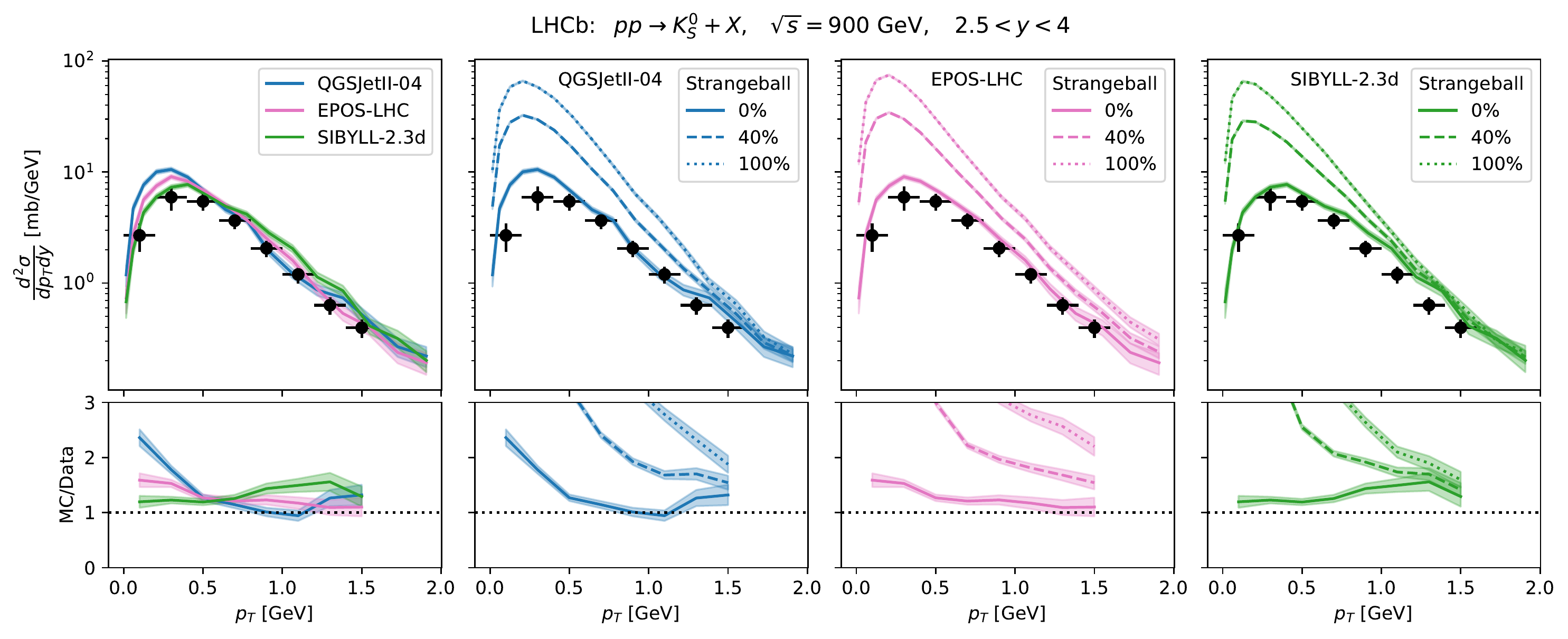}
    }
    \caption{\label{fig:LHCb} The $K_S^0$ $p_T$-spectrum from proton-proton collisions 
    at $\sqrt{s}~=~900$~GeV as measured by the LHCb detector \cite{2010LHCb} (data points) and retrieved from various models {with CRMC} (lines), for the rapidity range of 2.5 to 4. For clarity, we separately visualize the effect of 40\% and 100\% {strangeballs} for each of the hadronic interaction models. The bands correspond to $1\,\sigma$ Monte Carlo uncertainties.}
\end{figure}

The picture looks different when considering LHCb measurements of the $K^0_S$ $p_T$-spectrum produced by $\sqrt{s} = 900$ GeV proton-proton collisions \cite{2010LHCb}, as shown in figure~\ref{fig:LHCb}. Here, the introduction of 40\% or 100\% {strangeballs} significantly enhances the spectra as the more abundant neutral pions are converted to neutral kaons, breaking the reasonable agreement of the hadronic interaction models with the data. This seems to be a clear constraint on the {strangeball} model, especially given the relevant rapidity range for muon production (see figure~\ref{fig:rapidity}).
{However, two caveats need to be considered in the interpretation of this data: the energy and the species of the colliding particles.}

\begin{figure}[h]
    \centering
    \includegraphics[width=0.6\textwidth]{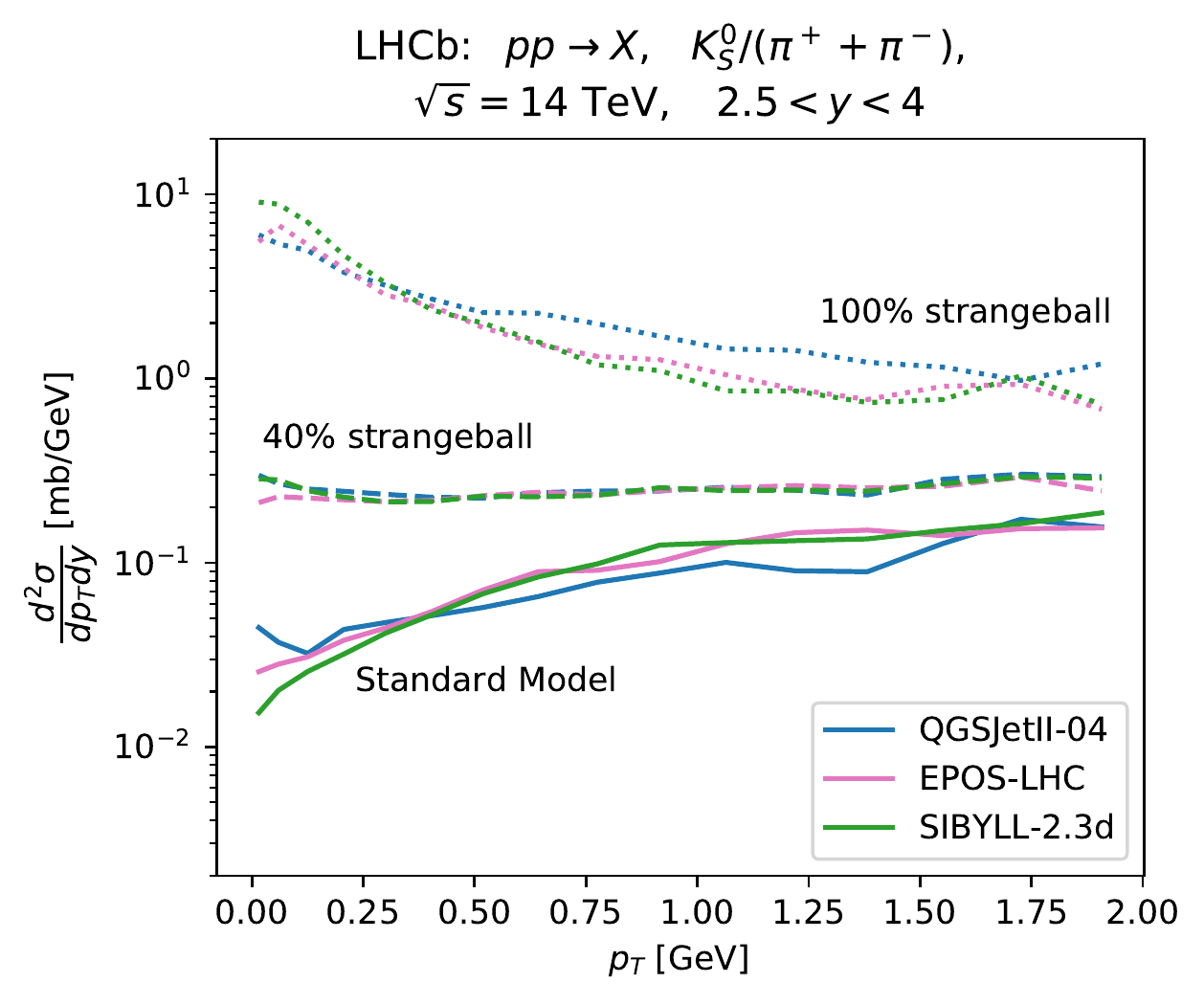}
    \caption{\label{fig:LHCbpredict} Ratio of $K_S^0$ to charged pion $p_T$-spectra from proton-proton collisions at $\sqrt{s}~=~14$~TeV{, computed with CRMC,} that could be measured by the LHCb detector to test the {strangeball} solution to the muon puzzle.}
\end{figure}

{For proton-proton collisions this data implies that strangeballs may only appear at higher energies}, requiring $E_\mathrm{min} > (900 \; \mathrm{GeV})^2/(2m_p) = 4\cdot10^{14}$ eV. Looking at the left plot in figure~\ref{fig:interpAugerbestfit} we then find solutions to the muon puzzle only at relatively low {values for $E_\mathrm{max}$ ($\lesssim 10^{18.5-19.5}$ eV), which competes} with the constraint arising from muon fluctuations{, requiring $E_\mathrm{max} \gtrsim 10^{19}$ eV. Exploiting this tension, the upcoming proton-proton LHC runs at $\sqrt{s}=14$ TeV provides an opportunity to rule out the formation of strangeballs \emph{from protons} as a solution to the muon puzzle.} We suggest in particular {the ratio of $K_S^0$} to the charged pion spectra as a clear indicator for {strangeball}-like effects, see figure~\ref{fig:LHCbpredict}.

{
It is important to emphasize that the previous discussion only concerns collisions between protons, and that the bulk of the interactions in air showers are those between pions and} air (see, e.g., figure~1 of ref.~\cite{2021Reininghaus}). {Nuclear effects seem to be negligible as} we found no difference in figure~\ref{fig:LHCbpredict} when considering proton-oxygen collisions at 10 TeV{. In contrast, having pions instead of protons as projectiles may play an important role.}
Unfortunately, it is experimentally challenging to produce charged pion beams for super-TeV center of mass collisions {and look for strangeballs directly. Until then we depend on the synergy between indirect measurements at collider and air shower experiments, and the further development of hadronic interaction models to make progress.

Here we would like to add to this synergy by making a qualitative argument on the consequence of having pions as projectiles for the viability of the strangeball model. Since pions are more compact than protons, the same new physics (e.g., strangeballs) should be induced at lower energies. Inversely, the previously derived constraint on $E_\mathrm{min}$ from $\sqrt{s}=900$ GeV LHCb data is relaxed for pions. The hadronic cascade in air showers is fueled by interactions between projectile partons with relatively large $x$ \cite{1997Hillas, 2021Reininghaus} and thus target partons with small $x$. At small $x \equiv E_\mathrm{parton}/E_\mathrm{hadron}$, the parton distribution function (PDF) is dominated by gluons, falling as $x\,g(x,Q^2) \propto x^\alpha$ with $-0.29 \leq \alpha \leq -0.14$ at $x=10^{-4}$ in various PDF sets \cite{2016Ball}. Estimating the volume difference between pions and protons based on the number of valence quarks, the gluon volume density of pions is a factor 3/2 that of protons. A proton-air interaction would achieve the same energy density at a factor $(3/2)^{1/\alpha}\approx 0.06-0.25$ in $x$, or $4-18$ in energy. This implies that for pion projectiles the constraint on $E_\mathrm{min}$ can be relaxed by about an order of magnitude, alleviating the tension with the muon fluctuations. Interestingly, this is not sufficient to avoid a constraint from measurements at $\sqrt{s}=14$ TeV $\Leftrightarrow E=10^{17}$ eV. Therefore, the 40\% strangeball deviation from the Standard Model shown in figure \ref{fig:LHCbpredict} is a concrete prediction of the strangeball model as a solution to the muon deficit.

}


\section{Conclusions}
\label{sec:conclusions}

Adopting the phenomenological modeling of ref.~\cite{AnchordoquiProceeding}, we exclude the production of fireballs in cosmic ray air showers as a resolution to the muon puzzle based on the fact that it introduces a tension between data on the average and fluctuations of $X_\mathrm{max}$. This can be traced back to the fireball property of producing a 
plasma state, which enhances the inelasticity (and multiplicity) and thereby accelerates the shower development. Instead, we find the effect on $X_\mathrm{max}$ to be negligible when ignoring the formation of a plasma (modeled by repeated \emph{in situ} collisions) and only considering enhancements to the strange quark content of the secondaries (modeled by swapping pions and kaons). This strangeness enhancement is found to be sufficient for resolving the muon puzzle at 10 EeV.

{To make a clear distinction with the original fireball model, we denote the \textit{strangeball} model as the model enhancing the strangeness without plasma formation. To test this strangeball model as a solution to the muon puzzle also at other energies,}
we extended the Heitler-Matthews model such that it can include variations in the fraction of energy remaining in the hadronic component. By fitting this to {strangeball}-extended \textsc{Conex} simulations we inferred the parameters relevant for the {strangeball} model. With a subsequent application to Auger data we find a spectrum of solutions to the muon puzzle, characterized by two extremes: a sudden introduction of {strangeballs} at $10^{17}$ eV, making up the first few interactions, or a gradual introduction starting already at low energies ($<10^{14}$ eV). Data on the relative muon fluctuations favors the latter extreme, implying that effects must be visible at energies accessible to collider experiments. At LHC and Tevatron energies, we find a solution to the muon puzzle to require 40\% of the interactions to be {strangeballs}, corresponding to a sizable $5-9$\% increase of the energy {retained in the hadronic cascade,} with respect to predictions from current hadronic interaction models.

From a comparison with {current} LHC measurements we cannot directly exclude this scenario. The theoretical uncertainty on the model predictions for the neutral pion spectra as measured by the LHCf detector exceeds the effect from a 40\% {strangeball}. While data from LHCb at $\sqrt{s} = 900$ GeV does exclude a {strangeball}, this constraint can be circumvented by only introducing {strangeballs} above $10^{14}$ eV. Note that this competes with the constraint from the muon fluctuations. {This tension is relaxed by an order of magnitude when recognizing that air showers are dominated by pion-air interactions.} We suggest a LHCb measurement of the ratio of $p_T$-spectra of $K_S^0$ to charged pions at $\sqrt{s} = 14$ TeV to be a powerful test for the {strangeball} scenario. 

As final remarks we would like to emphasize that still much can be won on the astroparticle side of the muon puzzle. For example, including data from the underground muon detector of the Pierre Auger Observatory \cite{2020AugerMuonUMD} --- and in its extension also data from other experiments (see, e.g., ref.~\cite{2019Dembinski}) --- in a similar analysis could further constrain the origin of the muon puzzle and provide more precise predictions for collider experiments. Along the same lines, also the reduction of the systematic and statistical uncertainties on respectively the average and relative fluctuations of the muon number will be beneficial. Finally, the concrete {strangeball} prediction of altering the strange particle content in air showers may affect more observables beyond the absolute muon number: the larger critical energy of kaons would give rise to a harder muon energy spectrum and a shallower muon production depth. The latter would actually improve the agreement with Auger data \cite{2014AugerMuonProdDepth}, motivating further studies in this direction as well.

It remains important to connect the particle and astroparticle physics communities as both are needed to solve the muon puzzle. The benefit to the astroparticle physics community is readily apparent, with precision determinations of the cosmic ray mass composition bringing us closer to {revealing} the cosmic ray origin. But the potential benefit to the particle physics community cannot be understated, considering the unique high-energy and ultra-high-energy laboratory provided by air showers. To sustainably drive this synergy we should work towards flexible (modular) and transparent hadronic interaction models. These would enable fundamental interpretations of astroparticle data in term of the underlying particle physics, as well as lay a solid foundation for studying the cosmos.

\acknowledgments

We thank Sergey Ostapchenko for lending us his expertise through critical comments and the ensuing discussions. We further thank Tanguy Pierog for his suggestions and help regarding the implementation of the fireball{/strangeball} model into \textsc{Conex}. The simulations were performed on the PHYSnet Compute Cloud of the University of Hamburg. Luis Anchordoqui is acknowledged for useful communications.
The work of M.V.G. was partially supported by the Bundesministerium f\"ur Bildung und Forschung, under contract 05H21GUCCA. J. M. and G.S. acknowledge support by the Bundesministerium f\"ur Bildung und Forschung, under grants 05A17GU1 and 05A20GU2.

\bibliographystyle{JHEP} 
\bibliography{references}

\providecommand{\href}[2]{#2}\begingroup\raggedright\begin{thebibliography}{10}

\bibitem{AugerSpectrum}
\scshape The Pierre Auger Collaboration, \emph{Measurement of the cosmic-ray
  energy spectrum above $2.5\ifmmode\times\else\texttimes\fi{}{10}^{18}\text{
  }\text{ }\mathrm{eV}$ using the Pierre Auger Observatory},
  \href{https://doi.org/10.1103/PhysRevD.102.062005}{\emph{Phys. Rev. D}
  {\bfseries 102} (2020) 062005}.

\bibitem{2014AugerXmaxmeasurement}
\scshape The Pierre Auger Collaboration, \emph{Depth of maximum of air-shower
  profiles at the Pierre Auger Observatory. I. Measurements at energies above
  $1{0}^{17.8}\text{ }\mathrm{eV}$},
  \href{https://doi.org/10.1103/PhysRevD.90.122005}{\emph{Phys. Rev. D}
  {\bfseries 90} (2014) 122005}.

\bibitem{2014AugerXmaxcomp}
\scshape The Pierre Auger Collaboration, \emph{Depth of maximum of air-shower
  profiles at the Pierre Auger Observatory. II. Composition implications},
  \href{https://doi.org/10.1103/PhysRevD.90.122006}{\emph{Phys. Rev. D}
  {\bfseries 90} (2014) 122006}.

\bibitem{2016AugerMixed}
\scshape The Pierre Auger Collaboration, \emph{Evidence for a mixed mass
  composition at the ‘ankle’ in the cosmic-ray spectrum},
  \href{https://doi.org/https://doi.org/10.1016/j.physletb.2016.09.039}{\emph{Physics
  Letters B} {\bfseries 762} (2016) 288}.

\bibitem{2017AugerInferencesTests}
\scshape The Pierre Auger Collaboration, \emph{Inferences on mass composition
  and tests of hadronic interactions from 0.3 to 100 EeV using the
  water-Cherenkov detectors of the Pierre Auger Observatory},
  \href{https://doi.org/10.1103/PhysRevD.96.122003}{\emph{Phys. Rev. D}
  {\bfseries 96} (2017) 122003}.

\bibitem{2019Dembinski}
{Dembinski, H.P.}, {Arteaga-Vel\'azquez, J.C.}, {Cazon, L.}, {Concei\c{c}\~ao,
  R.}, {Gonzalez, J.}, {Itow, Y.} et~al., \emph{Report on Tests and
  Measurements of Hadronic Interaction Properties with Air Showers},
  \href{https://doi.org/10.1051/epjconf/201921002004}{\emph{EPJ Web Conf.}
  {\bfseries 210} (2019) 02004}.

\bibitem{2021MuonPuzzle}
J.~Albrecht et~al., \emph{{The Muon Puzzle in cosmic-ray induced air showers
  and its connection to the Large Hadron Collider}},
  \href{https://doi.org/10.1007/s10509-022-04054-5}{\emph{Astrophys. Space
  Sci.} {\bfseries 367} (2022) 27}
  [\href{https://arxiv.org/abs/2105.06148}{{\ttfamily 2105.06148}}].

\bibitem{2021WHISP}
D.~Soldin, \emph{{Update on the Combined Analysis of Muon Measurements from
  Nine Air Shower Experiments}},  in \emph{Proceedings of 37th International
  Cosmic Ray Conference {\textemdash} PoS(ICRC2021)}, vol.~395, p.~349, 2021,
  \href{https://doi.org/10.22323/1.395.0349}{doi: 10.22323/1.395.0349}.

\bibitem{2010Ulrich}
R.~Ulrich, R.~Engel and M.~Unger, \emph{Hadronic multiparticle production at
  ultrahigh energies and extensive air showers},
  \href{https://doi.org/10.1103/PhysRevD.83.054026}{\emph{Phys. Rev. D}
  {\bfseries 83} (2011) 054026}.

\bibitem{2020Baur}
S.~Baur, H.~Dembinski, M.~Perlin, T.~Pierog, R.~Ulrich and K.~Werner,
  \emph{{Core-corona effect in hadron collisions and muon production in air
  showers}}, {\emph{arXiv e-prints} (2019) }
  [\href{https://arxiv.org/abs/1902.09265}{{\ttfamily 1902.09265}}].

\bibitem{2022AnchordoquiPiKswap}
L.A.~Anchordoqui, C.G.~Canal, F.~Kling, S.J.~Sciutto and J.F.~Soriano, \emph{An
  explanation of the muon puzzle of ultrahigh-energy cosmic rays and the role
  of the Forward Physics Facility for model improvement},
  \href{https://doi.org/https://doi.org/10.1016/j.jheap.2022.03.004}{\emph{Journal
  of High Energy Astrophysics} (2022) }.

\bibitem{JulienThesis}
J.~Manshanden, \emph{Ultra-High-Energy Cosmic Rays: A Fireball Model to resolve
  the Deficit of Muons in Simulations of Extensive Air Showers}, Ph.D. thesis,
  Universit{\"a}t Hamburg, 2021.
\newblock
  \href{https://ediss.sub.uni-hamburg.de/handle/ediss/9531}{https://ediss.sub.uni-hamburg.de/handle/ediss/9531}.

\bibitem{AnchordoquiFireball}
L.A.~Anchordoqui, H.~Goldberg and T.J.~Weiler, \emph{{Strange fireball as an
  explanation of the muon excess in Auger data}},
  \href{https://doi.org/10.1103/PhysRevD.95.063005}{\emph{Phys. Rev.}
  {\bfseries D95} (2017) 063005}
  [\href{https://arxiv.org/abs/1612.07328}{{\ttfamily 1612.07328}}].

\bibitem{2012Muniz}
J.~Alvarez-Muniz, L.~Cazon, R.~Conceicao, J.D.~de~Deus, C.~Pajares and
  M.~Pimenta, \emph{{Muon production and string percolation effects in cosmic
  rays at the highest energies}}, {\emph{arXiv e-prints} (2012) }
  [\href{https://arxiv.org/abs/1209.6474}{{\ttfamily 1209.6474}}].

\bibitem{2013Farrar}
{Farrar, Glennys R.} and {Allen, Jeffrey D.}, \emph{A new physical phenomenon
  in ultra-high energy collisions},
  \href{https://doi.org/10.1051/epjconf/20135307007}{\emph{EPJ Web of
  Conferences} {\bfseries 53} (2013) 07007}.

\bibitem{2019PierogICRC}
T.~Pierog, S.~Baur, H.~Dembinski, R.~Ulrich and K.~Werner, \emph{{Collective
  Hadronization and Air Showers: Can LHC Data Solve the Muon Puzzle ?}},  in
  \emph{Proceedings of 36th International Cosmic Ray Conference {\textemdash}
  PoS(ICRC2019)}, vol.~358, p.~387, 2019,
  \href{https://doi.org/10.22323/1.358.0387}{doi: 10.22323/1.358.0387}.

\bibitem{2021PierogICRC}
T.~Pierog, S.~Baur, H.~Dembinski, M.~Perlin, R.~Ulrich and K.~Werner,
  \emph{{When heavy ions meet cosmic rays: potential impact of QGP formation on
  the muon puzzle}},  in \emph{Proceedings of 37th International Cosmic Ray
  Conference {\textemdash} PoS(ICRC2021)}, vol.~395, p.~469, 2021,
  \href{https://doi.org/10.22323/1.395.0469}{doi: 10.22323/1.395.0469}.

\bibitem{2015AugerDetails}
\scshape The Pierre Auger Collaboration, \emph{The Pierre Auger Cosmic Ray
  Observatory},
  \href{https://doi.org/https://doi.org/10.1016/j.nima.2015.06.058}{\emph{Nucl.
  Instrum. Methods Phys. Res. A} {\bfseries 798} (2015) 172}.

\bibitem{1937Carlson}
J.F.~Carlson and J.R.~Oppenheimer, \emph{On Multiplicative Showers},
  \href{https://doi.org/10.1103/PhysRev.51.220}{\emph{Phys. Rev.} {\bfseries
  51} (1937) 220}.

\bibitem{1954Heitler}
W.~Heitler, \emph{The Quantum Theory of Radiation}, Oxford University Press
  (1954).

\bibitem{2005Matthews}
J.~Matthews, \emph{A Heitler model of extensive air showers},
  \href{https://doi.org/https://doi.org/10.1016/j.astropartphys.2004.09.003}{\emph{Astroparticle
  Physics} {\bfseries 22} (2005) 387}.

\bibitem{2009Andronic}
A.~Andronic, P.~Braun-Munzinger and J.~Stachel, \emph{{Thermal hadron
  production in relativistic nuclear collisions: The Hadron mass spectrum, the
  horn, and the QCD phase transition}},
  \href{https://doi.org/10.1016/j.physletb.2009.06.021}{\emph{Phys. Lett. B}
  {\bfseries 673} (2009) 142}
  [\href{https://arxiv.org/abs/0812.1186}{{\ttfamily 0812.1186}}].

\bibitem{2016Braun-Munzinger}
P.~Braun-Munzinger, V.~Koch, T.~Sch\"afer and J.~Stachel, \emph{{Properties of
  hot and dense matter from relativistic heavy ion collisions}},
  \href{https://doi.org/10.1016/j.physrep.2015.12.003}{\emph{Phys. Rept.}
  {\bfseries 621} (2016) 76}
  [\href{https://arxiv.org/abs/1510.00442}{{\ttfamily 1510.00442}}].

\bibitem{2016Dusling}
K.~Dusling, W.~Li and B.~Schenke, \emph{{Novel collective phenomena in
  high-energy proton\textendash{}proton and proton\textendash{}nucleus
  collisions}}, \href{https://doi.org/10.1142/S0218301316300022}{\emph{Int. J.
  Mod. Phys. E} {\bfseries 25} (2016) 1630002}
  [\href{https://arxiv.org/abs/1509.07939}{{\ttfamily 1509.07939}}].

\bibitem{2017Koch}
P.~Koch, B.~M\"uller and J.~Rafelski, \emph{{From Strangeness Enhancement to
  Quark\textendash{}Gluon Plasma Discovery}},
  \href{https://doi.org/10.1142/S0217751X17300241}{\emph{Int. J. Mod. Phys. A}
  {\bfseries 32} (2017) 1730024}
  [\href{https://arxiv.org/abs/1708.08115}{{\ttfamily 1708.08115}}].

\bibitem{2018Busza}
W.~Busza, K.~Rajagopal and W.~van~der Schee, \emph{{Heavy Ion Collisions: The
  Big Picture, and the Big Questions}},
  \href{https://doi.org/10.1146/annurev-nucl-101917-020852}{\emph{Ann. Rev.
  Nucl. Part. Sci.} {\bfseries 68} (2018) 339}
  [\href{https://arxiv.org/abs/1802.04801}{{\ttfamily 1802.04801}}].

\bibitem{2020Gazdzicki}
M.~Gazdzicki, M.~Gorenstein and P.~Seyboth, \emph{{Brief history of the search
  for critical structures in heavy-ion collisions}},
  \href{https://doi.org/10.5506/APhysPolB.51.1033}{\emph{Acta Phys. Polon. B}
  {\bfseries 51} (2020) 1033}
  [\href{https://arxiv.org/abs/2004.02255}{{\ttfamily 2004.02255}}].

\bibitem{2022ALICEQCDJourney}
\scshape ALICE, \emph{{The ALICE experiment -- A journey through QCD}},
  \href{https://arxiv.org/abs/2211.04384}{{\ttfamily 2211.04384}}.

\bibitem{2005BRAHMS}
\scshape BRAHMS, \emph{{Quark gluon plasma and color glass condensate at RHIC?
  The Perspective from the BRAHMS experiment}},
  \href{https://doi.org/10.1016/j.nuclphysa.2005.02.130}{\emph{Nucl. Phys. A}
  {\bfseries 757} (2005) 1}
  [\href{https://arxiv.org/abs/nucl-ex/0410020}{{\ttfamily nucl-ex/0410020}}].

\bibitem{2005PHENIX}
\scshape PHENIX, \emph{{Formation of dense partonic matter in relativistic
  nucleus-nucleus collisions at RHIC: Experimental evaluation by the PHENIX
  collaboration}},
  \href{https://doi.org/10.1016/j.nuclphysa.2005.03.086}{\emph{Nucl. Phys. A}
  {\bfseries 757} (2005) 184}
  [\href{https://arxiv.org/abs/nucl-ex/0410003}{{\ttfamily nucl-ex/0410003}}].

\bibitem{2005STAR}
\scshape STAR, \emph{{Experimental and theoretical challenges in the search for
  the quark gluon plasma: The STAR Collaboration's critical assessment of the
  evidence from RHIC collisions}},
  \href{https://doi.org/10.1016/j.nuclphysa.2005.03.085}{\emph{Nucl. Phys. A}
  {\bfseries 757} (2005) 102}
  [\href{https://arxiv.org/abs/nucl-ex/0501009}{{\ttfamily nucl-ex/0501009}}].

\bibitem{2008Becattini}
F.~Becattini and J.~Manninen, \emph{{Strangeness production from SPS to LHC}},
  \href{https://doi.org/10.1088/0954-3899/35/10/104013}{\emph{J. Phys. G}
  {\bfseries 35} (2008) 104013}
  [\href{https://arxiv.org/abs/0805.0098}{{\ttfamily 0805.0098}}].

\bibitem{2014ALICEpPb}
\scshape ALICE, \emph{{Multiplicity Dependence of Pion, Kaon, Proton and Lambda
  Production in p-Pb Collisions at $\sqrt{s_{NN}}$ = 5.02 TeV}},
  \href{https://doi.org/10.1016/j.physletb.2013.11.020}{\emph{Phys. Lett. B}
  {\bfseries 728} (2014) 25} [\href{https://arxiv.org/abs/1307.6796}{{\ttfamily
  1307.6796}}].

\bibitem{2016ALICEpPb}
\scshape ALICE, \emph{{Multi-strange baryon production in p-Pb collisions at
  $\sqrt{s_\mathbf{NN}}=5.02$ TeV}},
  \href{https://doi.org/10.1016/j.physletb.2016.05.027}{\emph{Phys. Lett. B}
  {\bfseries 758} (2016) 389}
  [\href{https://arxiv.org/abs/1512.07227}{{\ttfamily 1512.07227}}].

\bibitem{2017ALICEStrangenesspp}
\scshape ALICE, \emph{{Enhanced production of multi-strange hadrons in
  high-multiplicity proton-proton collisions}},
  \href{https://doi.org/10.1038/nphys4111}{\emph{Nature Phys.} {\bfseries 13}
  (2017) 535} [\href{https://arxiv.org/abs/1606.07424}{{\ttfamily
  1606.07424}}].

\bibitem{2017LaHurd}
D.~LaHurd and C.E.~Covault, \emph{{Exploring Potential Signatures of QGP in
  UHECR Ground Profiles}},
  \href{https://doi.org/10.1088/1475-7516/2018/11/007}{\emph{JCAP} {\bfseries
  11} (2018) 007} [\href{https://arxiv.org/abs/1707.01563}{{\ttfamily
  1707.01563}}].

\bibitem{AnchordoquiProceeding}
J.F.~Soriano, L.A.~Anchordoqui, T.C.~Paul and T.J.~Weiler, \emph{{Probing QCD
  approach to thermal equilibrium with ultrahigh energy cosmic rays}},
  \href{https://doi.org/10.22323/1.301.0342}{\emph{PoS} {\bfseries ICRC2017}
  (2018) 342} [\href{https://arxiv.org/abs/1811.07728}{{\ttfamily
  1811.07728}}].

\bibitem{2005Bozek}
P.~Bozek, \emph{{Size of the thermal source in relativistic heavy-ion
  collisions}}, {\emph{Acta Phys. Polon. B} {\bfseries 36} (2005) 3071}
  [\href{https://arxiv.org/abs/nucl-th/0506037}{{\ttfamily nucl-th/0506037}}].

\bibitem{2007Werner}
K.~Werner, \emph{{Core-corona separation in ultra-relativistic heavy ion
  collisions}},
  \href{https://doi.org/10.1103/PhysRevLett.98.152301}{\emph{Phys. Rev. Lett.}
  {\bfseries 98} (2007) 152301}
  [\href{https://arxiv.org/abs/0704.1270}{{\ttfamily 0704.1270}}].

\bibitem{2008Aichelin}
J.~Aichelin and K.~Werner, \emph{{Centrality Dependence of Strangeness
  Enhancement in Ultrarelativistic Heavy Ion Collisions: A Core-Corona
  Effect}}, \href{https://doi.org/10.1103/PhysRevC.79.064907}{\emph{Phys. Rev.
  C} {\bfseries 79} (2009) 064907}
  [\href{https://arxiv.org/abs/0810.4465}{{\ttfamily 0810.4465}}].

\bibitem{2009Becattini}
F.~Becattini and J.~Manninen, \emph{{Centrality dependence of strangeness
  production in heavy-ion collisions as a geometrical effect of core-corona
  superposition}},
  \href{https://doi.org/10.1016/j.physletb.2009.01.066}{\emph{Phys. Lett. B}
  {\bfseries 673} (2009) 19} [\href{https://arxiv.org/abs/0811.3766}{{\ttfamily
  0811.3766}}].

\bibitem{2015Block}
M.M.~Block, L.~Durand, P.~Ha and F.~Halzen, \emph{{Comprehensive fits to high
  energy data for $\sigma$, $\rho$, and $B$ and the asymptotic black-disk
  limit}}, \href{https://doi.org/10.1103/PhysRevD.92.114021}{\emph{Phys. Rev.
  D} {\bfseries 92} (2015) 114021}
  [\href{https://arxiv.org/abs/1511.02406}{{\ttfamily 1511.02406}}].

\bibitem{CONEX1}
T.~Pierog et~al., \emph{{First results of fast one-dimensional hybrid
  simulation of EAS using CONEX}},
  \href{https://doi.org/10.1016/j.nuclphysbps.2005.07.029}{\emph{Nucl. Phys. B
  Proc. Suppl.} {\bfseries 151} (2006) 159}
  [\href{https://arxiv.org/abs/astro-ph/0411260}{{\ttfamily
  astro-ph/0411260}}].

\bibitem{CONEX2}
T.~Bergmann, R.~Engel, D.~Heck, N.N.~Kalmykov, S.~Ostapchenko, T.~Pierog
  et~al., \emph{{One-dimensional Hybrid Approach to Extensive Air Shower
  Simulation}},
  \href{https://doi.org/10.1016/j.astropartphys.2006.08.005}{\emph{Astropart.
  Phys.} {\bfseries 26} (2007) 420}
  [\href{https://arxiv.org/abs/astro-ph/0606564}{{\ttfamily
  astro-ph/0606564}}].

\bibitem{CONEXxCorsika1}
T.~Pierog, R.~Engel and D.~Heck, \emph{3D Air Shower Simulations Using CONEX in
  CORSIKA},  in \emph{Proc. 31 st Int. Cosmic Ray Conf., Lodz (Poland), (2009)
  contr. 0425},
  \href{http://icrc2009.uni.lodz.pl/proc/pdf/icrc0425.pdf}{http://icrc2009.uni.lodz.pl/proc/pdf/icrc0425.pdf}.

\bibitem{CONEXxCorsika2}
T.~Pierog, R.~Engel, D.~Heck and R.~Ulrich, \emph{3D hybrid air shower
  simulation in CORSIKA},  in \emph{32nd Internat. Cosmic Ray Conf., Beijing,
  China, August 11-18, 2011. Vol. 2}, pp.~222--225, 2012,
  \href{https://doi.org/10.7529/ICRC2011/V02/1170}{doi:
  10.7529/ICRC2011/V02/1170}.

\bibitem{Corsika}
D.~Heck, J.~Knapp, J.~Capdevielle, G.~Schatz and T.~Thouw, \emph{CORSIKA: A
  Monte Carlo Code to Simulate Extensive Air Showers},
  \href{https://doi.org/10.5445/IR/270043064}{\emph{Forschungszentrum Karlsruhe
  Report FZKA} {\bfseries 6019} (1998) }.

\bibitem{2015AugerMuons}
\scshape The Pierre Auger Collaboration, \emph{{Muons in Air Showers at the
  Pierre Auger Observatory: Mean Number in Highly Inclined Events}},
  \href{https://doi.org/10.1103/PhysRevD.91.032003}{\emph{Phys. Rev. D}
  {\bfseries 91} (2015) 032003}
  [\href{https://arxiv.org/abs/1408.1421}{{\ttfamily 1408.1421}}].

\bibitem{2019ICRC}
\scshape The Pierre Auger Collaboration, \emph{{The Pierre Auger Observatory:
  Contributions to the 36th International Cosmic Ray Conference (ICRC 2019)}},
  {\emph{arXiv e-prints} (2019) }
  [\href{https://arxiv.org/abs/1909.09073}{{\ttfamily 1909.09073}}].

\bibitem{2011QGSJetII-04}
S.~Ostapchenko, \emph{Monte Carlo treatment of hadronic interactions in
  enhanced Pomeron scheme: QGSJET-II model},
  \href{https://doi.org/10.1103/PhysRevD.83.014018}{\emph{Phys. Rev. D}
  {\bfseries 83} (2011) 014018}.

\bibitem{2013QGSJetII-04}
S.~Ostapchenko, \emph{{QGSJET-II: physics, recent improvements, and results for
  air showers}}, \href{https://doi.org/10.1051/epjconf/20125202001}{\emph{EPJ
  Web Conf.} {\bfseries 52} (2013) 02001}.

\bibitem{EPOSLHC}
T.~Pierog, I.~Karpenko, J.M.~Katzy, E.~Yatsenko and K.~Werner, \emph{EPOS LHC:
  Test of collective hadronization with data measured at the CERN Large Hadron
  Collider}, \href{https://doi.org/10.1103/PhysRevC.92.034906}{\emph{Phys. Rev.
  C} {\bfseries 92} (2015) 034906}.

\bibitem{2020Sibyll23d_PRD}
F.~Riehn, R.~Engel, A.~Fedynitch, T.K.~Gaisser and T.~Stanev, \emph{Hadronic
  interaction model sibyll 2.3d and extensive air showers},
  \href{https://doi.org/10.1103/PhysRevD.102.063002}{\emph{Phys. Rev. D}
  {\bfseries 102} (2020) 063002}.

\bibitem{UrQMD1}
S.~Bass, M.~Belkacem, M.~Bleicher, M.~Brandstetter, L.~Bravina, C.~Ernst
  et~al., \emph{Microscopic models for ultrarelativistic heavy ion collisions},
  \href{https://doi.org/https://doi.org/10.1016/S0146-6410(98)00058-1}{\emph{Progress
  in Particle and Nuclear Physics} {\bfseries 41} (1998) 255}.

\bibitem{UrQMD2}
M.~Bleicher, E.~Zabrodin, C.~Spieles, S.A.~Bass, C.~Ernst, S.~Soff et~al.,
  \emph{Relativistic hadron-hadron collisions in the ultra-relativistic quantum
  molecular dynamics model},
  \href{https://doi.org/10.1088/0954-3899/25/9/308}{\emph{J. Phys. G: Nucl.
  Part. Phys.} {\bfseries 25} (1999) 1859}.

\bibitem{1960Fukui}
S.~Fukui, H.~Hasegawa, T.~Matano, I.~Miura, M.~Oda, K.~Suga et~al., \emph{{A
  Study on the Structure of the Extensive Air Shower}},
  \href{https://doi.org/10.1143/PTPS.16.1}{\emph{Progress of Theoretical
  Physics Supplement} {\bfseries 16} (1960) 1}.

\bibitem{2018Cazon}
L.~Cazon, R.~Conceição and F.~Riehn, \emph{Probing the energy spectrum of
  hadrons in proton air interactions at ultrahigh energies through the
  fluctuations of the muon content of extensive air showers},
  \href{https://doi.org/https://doi.org/10.1016/j.physletb.2018.07.026}{\emph{Physics
  Letters B} {\bfseries 784} (2018) 68}.

\bibitem{1984Biro}
T.~Biro, H.~Nielsen and J.~Knoll, \emph{Colour rope model for extreme
  relativistic heavy ion collisions},
  \href{https://doi.org/https://doi.org/10.1016/0550-3213(84)90441-3}{\emph{Nuclear
  Physics B} {\bfseries 245} (1984) 449}.

\bibitem{2014Bierlich}
C.~Bierlich, G.~Gustafson, L.~L\"onnblad and A.~Tarasov, \emph{{Effects of
  Overlapping Strings in pp Collisions}},
  \href{https://doi.org/10.1007/JHEP03(2015)148}{\emph{JHEP} {\bfseries 03}
  (2015) 148} [\href{https://arxiv.org/abs/1412.6259}{{\ttfamily 1412.6259}}].

\bibitem{2022BierlichJetMod}
C.~Bierlich, S.~Chakraborty, G.~Gustafson and L.~L\"onnblad, \emph{{Jet
  modifications from colour rope formation in dense systems of non-parallel
  strings}}, \href{https://doi.org/10.21468/SciPostPhys.13.2.023}{\emph{SciPost
  Phys.} {\bfseries 13} (2022) 023}
  [\href{https://arxiv.org/abs/2202.12783}{{\ttfamily 2202.12783}}].

\bibitem{2022BierlichStrange}
C.~Bierlich, S.~Chakraborty, G.~Gustafson and L.~L\"onnblad, \emph{{Strangeness
  enhancement across collision systems without a plasma}},
  \href{https://doi.org/10.1016/j.physletb.2022.137571}{\emph{Phys. Lett. B}
  {\bfseries 835} (2022) 137571}
  [\href{https://arxiv.org/abs/2205.11170}{{\ttfamily 2205.11170}}].

\bibitem{CRMC}
R.~Ulrich, T.~Pierog and C.~Baus, \emph{Cosmic Ray Monte Carlo Package, CRMC},
  Aug., 2021.
\newblock \href{https://doi.org/10.5281/zenodo.5270381}{doi:
  10.5281/zenodo.5270381}.

\bibitem{2019Sibyll23c_PRD}
A.~Fedynitch, F.~Riehn, R.~Engel, T.K.~Gaisser and T.~Stanev, \emph{Hadronic
  interaction model sibyll $2.3\mathrm{c}$ and inclusive lepton fluxes},
  \href{https://doi.org/10.1103/PhysRevD.100.103018}{\emph{Phys. Rev. D}
  {\bfseries 100} (2019) 103018}.

\bibitem{2017GrimmProceedings}
S.~Grimm, D.~Veberic and R.~Engel, \emph{{Heitler-Matthews model with
  leading-particle effect}},
  \href{https://doi.org/10.22323/1.301.0299}{\emph{PoS} {\bfseries ICRC2017}
  (2017) 299}.

\bibitem{2021AugerMuonFluc}
\scshape The Pierre Auger Collaboration, \emph{Measurement of the Fluctuations
  in the Number of Muons in Extensive Air Showers with the Pierre Auger
  Observatory},
  \href{https://doi.org/10.1103/PhysRevLett.126.152002}{\emph{Phys. Rev. Lett.}
  {\bfseries 126} (2021) 152002}.

\bibitem{1997Hillas}
A.~Hillas, \emph{Shower simulation: lessons from MOCCA},
  \href{https://doi.org/https://doi.org/10.1016/S0920-5632(96)00847-X}{\emph{Nuclear
  Physics B - Proceedings Supplements} {\bfseries 52} (1997) 29}.

\bibitem{2021Reininghaus}
M.~Reininghaus, R.~Ulrich and T.~Pierog, \emph{{Air shower genealogy for muon
  production}}, \href{https://doi.org/10.22323/1.395.0463}{\emph{PoS}
  {\bfseries ICRC2021} (2021) 463}
  [\href{https://arxiv.org/abs/2108.03266}{{\ttfamily 2108.03266}}].

\bibitem{2016LHCf}
\scshape LHCf Collaboration, \emph{Measurements of longitudinal and transverse
  momentum distributions for neutral pions in the forward-rapidity region with
  the LHCf detector},
  \href{https://doi.org/10.1103/PhysRevD.94.032007}{\emph{Phys. Rev. D}
  {\bfseries 94} (2016) 032007}.

\bibitem{2010LHCb}
\scshape LHCb Collaboration, \emph{{Prompt $K^0_s$ production in $pp$
  collisions at $\sqrt{s}=0.9~\rm{TeV}$}},
  \href{https://doi.org/10.1016/j.physletb.2010.08.055}{\emph{Phys. Lett. B}
  {\bfseries 693} (2010) 69} [\href{https://arxiv.org/abs/1008.3105}{{\ttfamily
  1008.3105}}].

\bibitem{2016Ball}
R.D.~Ball, E.R.~Nocera and J.~Rojo, \emph{{The asymptotic behaviour of parton
  distributions at small and large $x$}},
  \href{https://doi.org/10.1140/epjc/s10052-016-4240-4}{\emph{Eur. Phys. J. C}
  {\bfseries 76} (2016) 383}
  [\href{https://arxiv.org/abs/1604.00024}{{\ttfamily 1604.00024}}].

\bibitem{2020AugerMuonUMD}
\scshape The Pierre Auger Collaboration, \emph{{Direct measurement of the
  muonic content of extensive air showers between $2\times 10^{17}$ and
  $2\times 10^{18}~$eV at the Pierre Auger Observatory}},
  \href{https://doi.org/10.1140/epjc/s10052-020-8055-y}{\emph{Eur. Phys. J. C}
  {\bfseries 80} (2020) 751}.

\bibitem{2014AugerMuonProdDepth}
\scshape The Pierre Auger Collaboration, \emph{Muons in air showers at the
  Pierre Auger Observatory: Measurement of atmospheric production depth},
  \href{https://doi.org/10.1103/PhysRevD.90.012012}{\emph{Phys. Rev. D}
  {\bfseries 90} (2014) 012012}.

\end{thebibliography}\endgroup


\appendix
\section{Relative Muon Fluctuations}
\label{app:muonfluc}

The relative shower-to-shower fluctuation of the muon number $\sigma(N_\mu)/\langle N_\mu \rangle$ is mainly determined by the physics of the first interaction \cite{1960Fukui, 2018Cazon}: fluctuations from subsequent interactions average out. The Pierre Auger Collaboration measured the size of these fluctuations above 4 EeV \cite{2021AugerMuonFluc}, constraining the production of {strangeballs} above this energy. We consider the 
measurements presented at the ICRC in 2019 \cite{2019ICRC}.

\begin{figure}[h]
    \centering
    \includegraphics[width=1\textwidth]{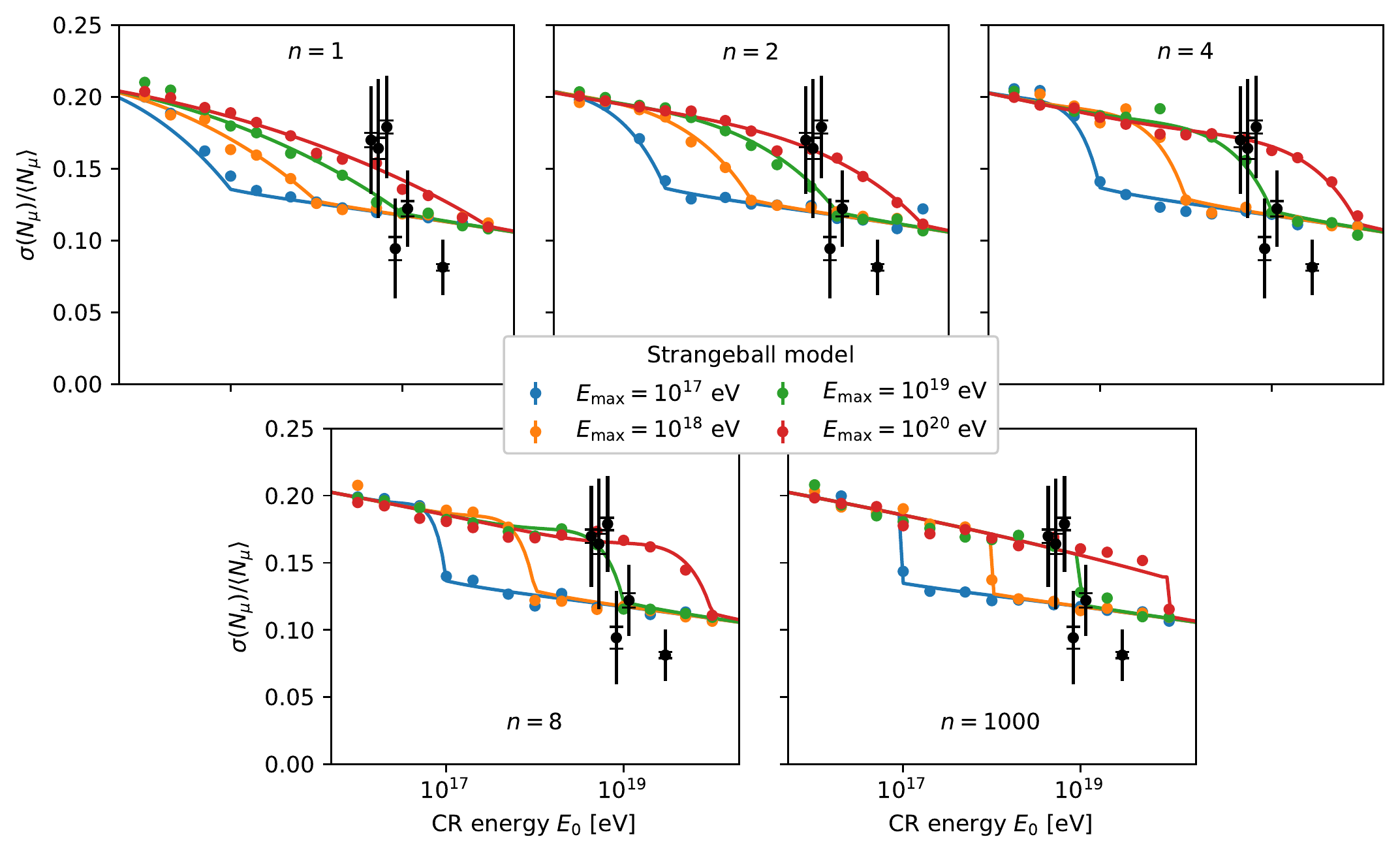}
    \caption{\label{fig:relfluc} Energy dependence of the relative muon fluctuations from {strangeball}-extended \textsc{Conex} simulations with EPOS-LHC (colored points) for various {strangeball} settings: $E_\mathrm{min} = 10^{15}$ eV is fixed, while $E_\mathrm{max}$ (colors) and $n$ (plots) vary. These simulations are fitted with our model (colored lines, eq.~(\ref{eq:relnmuintegral})), whose fit parameters are summarized in table~\ref{tab:flucfit}. The black data points are from the Pierre Auger Observatory as presented at the ICRC in 2019 \cite{2019ICRC}.}
\end{figure}

From our {strangeball}-extended \textsc{Conex} simulations we find that introducing the {strangeball} model reduces the relative muon fluctuations. This is visualized with proton primaries in figure~\ref{fig:relfluc}:
above $E_\mathrm{max}$ the fluctuations reach a slightly declining plateau between 0.15 and 0.10. 
For $n=1000$ and $E_0<E_\mathrm{max}$ one retrieves the Standard Model result, which corresponds to a similar plateau that is approximately 0.05 higher. The transition between the plateaus is determined by the {strangeball} settings through the {strangeball} probability $p(E)$ of eq.~(\ref{eq:probfireball}). A suppression of the relative muon fluctuations may be a general feature of any model that increases the hadronic energy fraction. We will get back to this at the end of the appendix.

We exclude the {strangeball} settings for which the predictions for proton CR primaries  
fall more than 1$\sigma$ (combined systematic and statistical uncertainties) below the data points from the Pierre Auger Observatory. Note that these constraints are actually more stringent since the $X_\mathrm{max}$-inferred composition requires the data to be below the proton prediction 
(see, e.g., figure~2 of ref.~\cite{2021AugerMuonFluc}). The constraints turn out to be mostly on $E_\mathrm{max}$, with that from the first (third) data point excluding the region to the left of the dashed (double-dot dashed) lines in figure~\ref{fig:interpAugerSCAN}.


To derive these constraints we constructed a model describing the energy dependence of the relative muon fluctuations, as described below.

\subsection{Modeling Relative Muon Fluctuations}

We follow a similar approach as in section~\ref{sec:HeitlerMatthews}, starting from the principles of the Heitler-Matthews model.

With the critical energy fixing the total number of particles $N_\mathrm{tot}$ and a generational picture of air showers, the muon number $N_\mu$ can be decomposed into a product of hadronic energy fractions $r_i$ at generation $i$ from the first generation until the critical generation $k_c$:
\begin{align}
    N_\mu  = N_\mathrm{tot} \prod_{i=1}^{k_c} r_i \;.
\end{align}

Fluctuations then arise from variations of these hadronic energy fractions and from variations of the total number of generations. For our modeling, we fix the total number of generations to $k_c$, effectively ignoring shower-to-shower fluctuations of the multiplicity. This is expected to be a subdominant effect since the value of $\beta$ (see eq.~(\ref{eq:NmuHM})) is close to 1 when inferred from Standard Model simulations. See ref.~\cite{2018Cazon} and in particular the discussion around eq.~(4) of this reference for more details.

A generational picture of air showers implies that the hadronic energy fraction at each generation is independent of the next. Therefore, the variance of the muon number can be decomposed as
\begin{align}
        \mathrm{var}(N_\mu) = \mathrm{var}\left(N_\mathrm{tot} \prod_{i=1}^{k_c} r_i \right) = N_\mathrm{tot}^2 \prod_{i=1}^{k_c} \left[ \sigma^2(r_i) + \langle r_i \rangle^2 \right] - N_\mathrm{tot}^2 \prod_{i=1}^{k_c} \langle r_i \rangle^2 \;,
\end{align}
where $\sigma^2(r_i)$ and $\langle r_i \rangle$ are the variance and average of the hadronic energy fraction at generation $i$, respectively. The relative muon fluctuations can then be written as
\begin{align}
        \frac{\sigma(N_\mu)}{\langle N_\mu \rangle} = \sqrt{\prod_{i=1}^{k_c} \left[1+ \left(\frac{\sigma(r_i)}{\langle r_i \rangle}\right)^2\right] - 1} \;. \label{eq:relnmu}
\end{align}
Here $r_i$ is the hadronic energy fraction of an entire generation and its fluctuations are suppressed by the number of particles $N_i$ in that generation: $\sigma(r_i) = \sigma(r_\mathrm{eff}) / \sqrt{N_i}$. This explicitly shows the dominance of the first interaction.

The effective hadronic energy fraction of an interaction $r_\mathrm{eff}$ is given by a superposition of {strangeball} and Standard Model interactions (eq.~(\ref{eq:refffb})), which can be propagated to the 
(squared) relative fluctuations as
\begin{align}
        \frac{\sigma^2(r_\mathrm{eff})}{\langle r_\mathrm{eff} \rangle^2} = \frac{(1-p) \; \sigma^2(r_\mathrm{SM}) + p \; \sigma^2(r_\mathrm{sb})}{\left[(1-p) \; \langle r_\mathrm{SM} \rangle + p \; \langle r_\mathrm{sb} \rangle\right]^2} + \frac{p(1-p) \left[ \langle r_\mathrm{sb} \rangle - \langle r_\mathrm{SM} \rangle \right]^2 }{\left[(1-p) \; \langle r_\mathrm{SM} \rangle + p \; \langle r_\mathrm{sb} \rangle\right]^2} \;,
    \label{eq:relnmuinteraction}
\end{align}
where we suppressed the energy dependence of $p \equiv p(E)$ for readability.

From a comparison with \textsc{Conex} simulations it turns out that a continuous extension of the product in eq.~(\ref{eq:relnmu}) provides a better description (see ref.~\cite{JulienThesis} for more details):
\begin{align}
        \prod_{i=1}^{k_c} \left[1 + \left(\frac{\sigma(r_i)}{\langle r_i \rangle}\right)^2 \right] \to \exp \left\{ \int_0^{k_c} \log \left[1 + \left(\frac{\sigma(r_i)}{\langle r_i \rangle}\right)^2 \right] dk + \Delta_\mathrm{corr} \right\}\;.
\end{align}
Similar to eq.~(\ref{eq:muonderiv}) we have a correction factor,
\begin{align}
    \Delta_\mathrm{corr} = \frac{1}{2} \log \left[1 + \left(\left.\frac{\sigma(r_i)}{\langle r_i \rangle}\right|_{k=0}\right)^2 \right] \;, \label{eq:relnmucorr}
\end{align}
corresponding to half a generation of the integrand, with the ratio of moments evaluated at the first interaction. Physically, this continuous representation allows us to include particles with energies `between' the generations and thereby partly take into account the elasticity of interactions in real air showers.

We nevertheless stick to a generational picture with synchronized shower branches and the energy equally divided over all secondaries. As in section~\ref{sec:HeitlerMatthews}, this allows us to use the multiplicity to relate the generation $k$ to the energy $E$ of the secondaries, and the total number of particles is simply $N=E_0/E$.

Putting everything together we have
\begin{align}
    \frac{\sigma(N_\mu)}{\langle N_\mu \rangle} = \sqrt{ \exp \left[ \int_{\log E_0}^{\log E_c} \log \left( 1+\frac{\sigma^2(r_\mathrm{eff})}{\langle r_\mathrm{eff} \rangle^2} \frac{E}{E_0} \right) \frac{dk}{d\log E}\; d\log E + \frac{1}{2} \log \left(1 + \left.\frac{\sigma^2(r_\mathrm{eff})}{\langle r_\mathrm{eff} \rangle^2}\right|_{E=E_0} \right) \right] - 1}\;, \label{eq:relnmuintegral}
\end{align}
with $\sigma^2(r_\mathrm{eff})/\langle r_\mathrm{eff} \rangle^2$ given in eq.~(\ref{eq:relnmuinteraction}) and the Jacobian for a power-law multiplicity (eq.~(\ref{eq:PLmult})) given by
\begin{align}
    \frac{dk}{d\log E} = \frac{1}{\log(1-b)} \frac{1}{\log \left(n_\mathrm{scale}^{1/b} E/\mathrm{GeV} \right)} \;.
\end{align}

These two terms significantly complicate the energy dependence of the integrand, preventing an analytic evaluation of the integral. Considering eq.~(\ref{eq:relnmuinteraction}), not only the {strangeball} probability $p(E)$ (eq.~(\ref{eq:probfireball})), but also $\langle r_\mathrm{SM} \rangle$, $\langle r_\mathrm{sb} \rangle$, $\sigma(r_\mathrm{SM})$, and $\sigma(r_\mathrm{sb})$ may be energy dependent. It turns out that the declining plateau features in figure~\ref{fig:relfluc} cannot be reproduced without this additional energy dependence, and thus we parametrize the variances as $\sigma^2(x) = \alpha - \beta \log_{10} (E/\mathrm{GeV})$, with $x\in\{r_\mathrm{SM}, r_\mathrm{sb}\}$.

With this we obtain a good fit to the \textsc{Conex} simulations as shown in figure~\ref{fig:relfluc} for EPOS-LHC. Similarly good fits are obtained for \textsc{QGSJetII-04} and \textsc{Sibyll-2.3d}, with all parameters summarized in table~\ref{tab:flucfit}. Note that these parameters differ from those listed in table~\ref{tab:paramestimates}, indicating that our picture is still a simplification of real air showers. We also found a good agreement with \textsc{Conex} simulations at $E_\mathrm{min} = 10^{14}$ eV and $10^{16}$ eV, justifying the interpolation in also this {strangeball} setting.


We end with a comment on the suppression of the relative muon fluctuations. At energies sufficiently above $E_\mathrm{max}$, the first part of the shower consists of {strangeballs} and thus $\sigma(N_\mu)/\langle N_\mu \rangle$ is an 
accumulation 
of $\sigma(r_\mathrm{sb})/\langle r_\mathrm{sb} \rangle$ from the first few interactions. The same holds for the Standard Model at energies below $E_\mathrm{min}$ (or below $E_\mathrm{max}$ with high $n$). The energy dependencies of $\sigma(r_\mathrm{sb})/\langle r_\mathrm{sb} \rangle$ and $\sigma(r_\mathrm{SM})/\langle r_\mathrm{SM} \rangle$  
thus form the declining plateau features, whose heights are determined by the sizes of these relative fluctuations. 
Since the hadronic energy fraction is bound from above by 1 (which limits the fluctuations), and from CRMC simulations we have $0.5 < \langle r_\mathrm{SM} \rangle < \langle r_\mathrm{sb} \rangle$ (table~\ref{tab:paramestimates}), we can expect $\sigma(r_\mathrm{sb})/\langle r_\mathrm{sb} \rangle < \sigma(r_\mathrm{SM})/\langle r_\mathrm{SM} \rangle$. 
This propagates to a plateau with a lower height 
for the {strangeball} model. The same argument holds for any model that increases the average hadronic energy fraction, with the exception of fine-tuned scenarios invoking a simultaneous, disproportional increase of the probability of events with low hadronic energy fractions. 

\newcommand*{\UniWidthh}[1]{{#1}}
\begin{table}[h]
\centering
\caption{Fit parameters for modeling the relative muon fluctuations of {strangeball}-extended \textsc{Conex} simulations with eq.~(\ref{eq:relnmuintegral}). The variances of the hadronic energy fractions are parametrized as $\sigma^2 = \alpha - \beta \log_\mathrm{10} (E/\mathrm{GeV})$.}
\label{tab:flucfit}
\medskip
\begingroup
\renewcommand{\arraystretch}{1.1} 
\begin{tabular}{c|cccccccc|}
\cline{2-9}
                                  & \multirow{2}{*}{\UniWidthh{$\langle r_\mathrm{SM} \rangle$}} & \multirow{2}{*}{\UniWidthh{$\langle r_\mathrm{sb} \rangle$}} & \multicolumn{2}{c}{$\sigma^2(r_\mathrm{SM})$} & \multicolumn{2}{c}{$\sigma^2(r_\mathrm{sb})$} & \multirow{2}{*}{\UniWidthh{$b$}} & \multirow{2}{*}{\UniWidthh{$n_\mathrm{scale}$}} \\ \cline{4-7}
                                  &                                                  &                                                  & \multicolumn{1}{|c}{\UniWidthh{$\alpha$}}   & \multicolumn{1}{c|}{\UniWidthh{$\beta$}}   & \UniWidthh{$\alpha$}   & \multicolumn{1}{c|}{\UniWidthh{$\beta$}}   &                      &                                     \\ \hline
\multicolumn{1}{|c|}{QGSJetII-04} & 0.669                                           & 0.840                                           & 0.100    & 0.00196                       & 0.0743    & 0.00140                       & 0.00184             & 51.29                               \\
\multicolumn{1}{|c|}{EPOS-LHC}    & 0.504                                           & 0.643                                           & 0.0621    & 0.00114                       & 0.0428    & 0.000719                      & 0.190               & 2722                                \\
\multicolumn{1}{|c|}{Sibyll-2.3d} & 0.531                                           & 0.658                                           & 0.0663    & 0.00117                       & 0.0527    & 0.000897                      & 0.177               & 2429                                \\ \hline
\end{tabular}
\endgroup
\end{table}


\clearpage

\end{document}